\documentclass[notitlepage,a4paper,aps,prd,onecolumn,superscriptaddress,nofootinbib,groupedaddress,longbibliography]{revtex4-2}
\usepackage{amsmath}
\usepackage{amsfonts}
\usepackage{amssymb}
\usepackage{accents}
\usepackage[utf8]{inputenc}
\usepackage[T1]{fontenc}
\usepackage[colorlinks=true]{hyperref}
\usepackage{tikz}
\usepackage[all]{hypcap}
\usepackage{enumerate}
\hypersetup{colorlinks=true}
\usetikzlibrary{arrows.meta,external}

\newcommand{\order}[2]{\accentset{#2}{#1}}
\newcommand{\lc}[1]{\accentset{\circ}{#1}}
\newcommand{\dd}{\mathrm{d}}
\newcommand{\vek}[1]{\underline{#1}}
\newcommand{\mat}[1]{\underline{\underline{#1}}}

\begin{document}

\title{Post-Newtonian limit of generalized scalar-teleparallel theories of gravity}

\author{Manuel Hohmann}
\email{manuel.hohmann@ut.ee}
\affiliation{Laboratory of Theoretical Physics, Institute of Physics, University of Tartu, W. Ostwaldi 1, 50411 Tartu, Estonia}

\author{Ulbossyn Ualikhanova}
\email{ulbossyn.ualikhanova@gmail.com}
\affiliation{Department of General and Theoretical Physics, L.~N.~Gumilyov Eurasian National University, Satpayev Str. 2, 010008 Astana, Kazakhstan}

\begin{abstract}
We propose a general class of scalar-teleparallel theories, which are based on a scalar field which is coupled to a flat connection with torsion and nonmetricity, and study its post-Newtonian limit using the parametrized post-Newtonian formalism. We find that among this class there are theories whose post-Newtonian limit fully agrees with general relativity; for others only the parameters \(\beta\) and \(\gamma\) deviate from their general relativity values \(\beta = \gamma = 1\), while all other parameters remain the same, thus preserving total momentum conservation, local Lorentz invariance and local position invariance; finally, we also find theories whose post-Newtonian limit is pathological. Our main result is a full classification of the proposed theories into these different cases. We apply our findings to a number of simpler classes of theories and show that for these a subset of the aforementioned cases can be found.
\end{abstract}

\maketitle


\section{Introduction}\label{sec:intro}
General relativity (GR) has been highly successful in describing gravity as the spacetime curvature. It has been extensively tested and confirmed in a wide range of observations, including the predictions of the bending of light, the precession of Mercury's orbit, and the existence of black holes and gravitational waves~\cite{Gillessen:2008qv,LIGOScientific:2016aoc,EventHorizonTelescope:2019dse,EventHorizonTelescope:2022wkp}. However successful in explaining these observations, the failure to describe GR as a quantum field theory alongside the other fundamental forces, and a number of unexplained observations in cosmology~\cite{Planck:2018vyg,DiValentino:2021izs}, have raised fundamental questions to which no conclusive answer has been found thus far. In order to find suitable models that can effectively explain the phenomena of dark energy, dark matter, and inflation, or are more accessible to quantization, numerous researchers have extended their investigations beyond the realm of GR~\cite{Nojiri:2006ri,Nojiri:2010wj,Capozziello:2010zz,Clifton:2011jh,Nojiri:2017ncd,Bull:2015stt,Heisenberg:2018vsk,CANTATA:2021ktz,Capozziello:2022zzh,Odintsov:2023weg}.

A geometric framework used to describe gravity which is currently under active development gives rise to the family of teleparallel gravity theories. While this family is most often described as two third of a trinity~\cite{BeltranJimenez:2019tjy}, in which a central role is played by a flat connection which exhibits either only torsion or only nonmetricity instead of curvature, commonly known as metric teleparallel gravity and symmetric teleparallel gravity, it is easily extended by allowing for a flat connection which allows for both torsion and nonmetricity, and constitutes the foundation of the so far less explored class of general teleparallel gravity theories~\cite{BeltranJimenez:2019odq,Hohmann:2022mlc,Heisenberg:2022mbo,Heisenberg:2023tho,Heisenberg:2023wgk}. Among this class, one finds the general teleparallel equivalent of general relativity (GTEGR), which may serve as a starting point for the construction of modified theories, e.g., by modifying its action or introducing further fundamental fields besides the metric and the flat connection. The most simple modification of the latter type is the addition of a scalar field, which then leads to a class of theories which can be subsumed under the name of scalar-teleparallel gravity theories~\cite{Hohmann:2022mlc,Heisenberg:2022mbo}. While being simple in their mathematical foundation and construction, scalar-teleparallel gravity offers a rich class of theories, which invite for further studies to address the aforementioned open questions in gravity theory.

While the primary motivation behind the construction of new gravity theories is to address the aforementioned open questions which are unanswered by GR, it is crucial for such a theory to also align with observations which are well explained by GR, e.g., the solar system, orbiting pulsars, and laboratory experiments. To achieve this, the parametrized post-Newtonian (PPN) formalism has been widely employed as a framework for deriving local-scale phenomenology~\cite{Will:1993ns,Will:2014kxa,Will:2018bme}. The PPN formalism characterizes gravity theories through ten parameters, which have been measured with great precision in various experiments~\cite{Bertotti:2003rm,Fienga:2011qh,Fienga:2014bvy,Verma:2013ata}. Due to its versatility and the abundance of available observations, the PPN formalism has become a valuable tool for assessing the viability of gravity theories, including teleparallel theories within the geometric trinity and theories which contain additional scalar fields~\cite{Nordtvedt:1970uv,Gladchenko:1990nw,Olmo:2005hc,Clifton:2010hz,Perivolaropoulos:2009ak,Scharer:2014kya,Hohmann:2013oca,Li:2013oef,Hohmann:2015kra,Hohmann:2016yfd,Hohmann:2017uxe,Sadjadi:2016kwj,Hohmann:2017uxe,Hohmann:2019qgo,Ualikhanova:2019ygl,Emtsova:2019qsl,Flathmann:2019khc,Flathmann:2021khc}.

The first goal of this article is to develop the PPN formalism for general teleparallel gravity theories, whose fundamental field is a general flat connection next to the metric. For this purpose, we introduce a suitable perturbative expansion of the connection in velocity orders, which is a key ingredient to the PPN formalism. Using this formalism, we can address our second objective to apply this general method to determine the post-Newtonian limit of a broad class of scalar-teleparallel gravity theories. This class has been chosen sufficiently general such that it encompasses a wide range of theories which may be studied as candidates to explain the so far unexplained observations in cosmology and other open questions in gravity theory. We then determine which of these theories possess a post-Newtonian limit which is compatible with observations and thus constitute viable candidate theories.

The structure of the current paper can be summarized as follows. In section~\ref{sec:stp} we briefly review the foundations of general teleparallel gravity and introduce the class of scalar-teleparallel gravity theories we study in this article. In section~\ref{sec:ppn} we review the basic ingredients of the post-Newtonian formalism, and show how it can be adapted to the field variables relevant for scalar-teleparallel gravity. We employ this formalism in order to solve the field equations for a general post-Newtonian matter distribution in section~\ref{sec:solution}. From this solution we obtain the post-Newtonian metric and PPN parameters in section~\ref{sec:ppnpar}. We summarize our findings to obtain a complete classification of the investigated scalar-teleparallel gravity theories in section~\ref{sec:classes}. Finally, in section~\ref{sec:examples} we discuss a number of specific examples. We end with a conclusion in section~\ref{sec:conclusion}.

\section{Scalar-teleparallel gravity}\label{sec:stp}
The class of theories we discuss in this article belongs to the geometric framework of general teleparallel gravity~\cite{BeltranJimenez:2019odq,Hohmann:2022mlc}, in which the dynamical fields are given by a metric \(g_{\mu\nu}\) and an affine connection with coefficients \(\Gamma^{\mu}{}_{\nu\rho}\), which is imposed to be flat,
\begin{equation}\label{eq:curvature}
R^{\mu}{}_{\nu\rho\sigma} = \partial_{\rho}\Gamma^{\mu}{}_{\nu\sigma} - \partial_{\sigma}\Gamma^{\mu}{}_{\nu\rho} + \Gamma^{\mu}{}_{\tau\rho}\Gamma^{\tau}{}_{\nu\sigma} - \Gamma^{\mu}{}_{\tau\sigma}\Gamma^{\tau}{}_{\nu\rho} \equiv 0\,.
\end{equation}
These two fundamental fields define the torsion
\begin{equation}\label{eq:torsion}
T^{\mu}{}_{\nu\rho} = \Gamma^{\mu}{}_{\rho\nu} - \Gamma^{\mu}{}_{\nu\rho}\,,
\end{equation}
as well as the nonmetricity
\begin{equation}\label{eq:nonmetricity}
Q_{\mu\nu\rho} = \nabla_{\mu}g_{\nu\rho} = \partial_{\mu}g_{\nu\rho} - \Gamma^{\sigma}{}_{\nu\mu}g_{\sigma\rho} - \Gamma^{\sigma}{}_{\rho\mu}g_{\nu\sigma}\,.
\end{equation}
Further, one defines the contortion
\begin{equation}\label{eq:contortion}
K^{\mu}{}_{\nu\rho} = \frac{1}{2}\left(T_{\nu}{}^{\mu}{}_{\rho} + T_{\rho}{}^{\mu}{}_{\nu} - T^{\mu}{}_{\nu\rho}\right)\,,
\end{equation}
and the disformation
\begin{equation}\label{eq:disformation}
L^{\mu}{}_{\nu\rho} = \frac{1}{2}\left(Q^{\mu}{}_{\nu\rho} - Q_{\nu}{}^{\mu}{}_{\rho} - Q_{\rho}{}^{\mu}{}_{\nu}\right)\,.
\end{equation}
These allow writing the difference between the coefficients of the teleparallel connection and those of the Levi-Civita connection, i.e., the Christoffel symbols
\begin{equation}\label{eq:levicivita}
\lc{\Gamma}^{\mu}{}_{\nu\rho} = \frac{1}{2}g^{\mu\sigma}\left(\partial_{\nu}g_{\sigma\rho} + \partial_{\rho}g_{\nu\sigma} - \partial_{\sigma}g_{\nu\rho}\right)\,,
\end{equation}
which we denote with a circle on top in order to distinguish them from the teleparallel connection, as
\begin{equation}\label{eq:conndec}
\Gamma^{\mu}{}_{\nu\rho} - \lc{\Gamma}^{\mu}{}_{\nu\rho} = M^{\mu}{}_{\nu\rho} = K^{\mu}{}_{\nu\rho} + L^{\mu}{}_{\nu\rho}\,.
\end{equation}
Here, \(M^{\mu}{}_{\nu\rho}\) is called the distortion. As a consequence of this decomposition, as well as the flatness condition~\eqref{eq:curvature}, we can write the curvature of the Levi-Civita connection as
\begin{equation}\label{eq:curvdec}
\lc{R}^{\mu}{}_{\nu\rho\sigma} = -\lc{\nabla}_{\rho}M^{\mu}{}_{\nu\sigma} + \lc{\nabla}_{\sigma}M^{\mu}{}_{\nu\rho} - M^{\mu}{}_{\tau\rho}M^{\tau}{}_{\nu\sigma} + M^{\mu}{}_{\tau\sigma}M^{\tau}{}_{\nu\rho}\,.
\end{equation}
In particular, it follows that the Ricci scalar \(\lc{R}\) is given by
\begin{equation}\label{eq:riccisplit}
\lc{R} = -G + B\,,
\end{equation}
where we defined the terms
\begin{equation}\label{eq:gtegrbndterms}
G = 2M^{\mu}{}_{\rho[\mu}M^{\rho\nu}{}_{\nu]}\,, \quad
B = 2\lc{\nabla}_{\mu}M^{[\nu\mu]}{}_{\nu}\,.
\end{equation}
Making use of this decomposition in the Einstein-Hilbert action
\begin{equation}\label{eq:einsteinhilbert}
S_{\text{EH}} = \frac{1}{2\kappa^2}\int_M\lc{R}\sqrt{-g}\dd^4x = \frac{1}{2\kappa^2}\int_M(-G + B)\sqrt{-g}\dd^4x\,,
\end{equation}
one sees that \(B\) becomes a boundary term, and so it does not contribute to the field equations. Hence, the latter are unchanged if it is omitted, leading to the action of the general teleparallel equivalent of general relativity (GTEGR) given by~\cite{Hohmann:2022mlc}
\begin{equation}\label{eq:gtegraction}
S_{\text{GTEGR}} = -\frac{1}{2\kappa^2}\int_MG\sqrt{-g}\dd^4x\,.
\end{equation}
The GTEGR action provides a starting point for different classes of modified gravity theories. In this article, we discuss a general class of scalar-teleparallel gravity theories, which employ a scalar field \(\phi\) as an additional field variable~\cite{Hohmann:2022mlc,Heisenberg:2022mbo}. In order to keep our discussion as general as possible, so that our results will be applicable to a wide range of theories, we consider a general form of the action given by
\begin{equation}\label{eq:staction}
S_{\text{ST}} = -\frac{1}{2\kappa^2}\int_ML(G, X, U, V, W, \phi)\sqrt{-g}\dd^4x\,,
\end{equation}
where we defined the terms
\begin{equation}
X = -\frac{1}{2}g^{\mu\nu}\phi_{,\mu}\phi_{,\nu}\,, \quad
U = T_{\mu}{}^{\mu\nu}\phi_{,\nu}\,, \quad
V = Q^{\nu\mu}{}_{\mu}\phi_{,\nu}\,, \quad
W = Q_{\mu}{}^{\mu\nu}\phi_{,\nu}\,.
\end{equation}
Its variation can be written as
\begin{equation}
\delta S_{\text{ST}} = -\frac{1}{2\kappa^2}\int_M\left[L_G\delta G + L_X\delta X + L_U\delta U + L_V\delta V + L_W\delta W + L_{\phi}\delta\phi + \frac{1}{2}Lg^{\mu\nu}\delta g_{\mu\nu}\right]\sqrt{-g}\dd^4x\,,
\end{equation}
where we have denoted derivatives of \(L\) with respect to its arguments with subscripts, omitted the arguments for brevity, and made use of the variations
\begin{subequations}
\begin{align}
\delta G &= \left(M^{\rho\sigma(\mu}M_{\sigma}{}^{\nu)}{}_{\rho} - M^{\rho(\mu\nu)}M^{\sigma}{}_{\rho\sigma}\right)\delta g_{\mu\nu} + \left(M^{\rho(\mu\nu)} - M^{\sigma(\mu}{}_{\sigma}g^{\nu)\rho} - M^{[\rho\sigma]}{}_{\sigma}g^{\mu\nu}\right)\lc{\nabla}_{\rho}\delta g_{\mu\nu}\nonumber\\
&\phantom{=}+ \left(M^{\nu\sigma}{}_{\sigma}\delta_{\mu}^{\rho} + M^{\sigma}{}_{\mu\sigma}g^{\nu\rho} - M^{\nu\rho}{}_{\mu} - M^{\rho}{}_{\mu}{}^{\nu}\right)\delta\Gamma^{\mu}{}_{\nu\rho}\,,\\
\delta X &= \frac{1}{2}\lc{\nabla}^{\mu}\phi\lc{\nabla}^{\nu}\phi\delta g_{\mu\nu} - \lc{\nabla}^{\mu}\phi\lc{\nabla}_{\mu}\delta\phi\,,\\
\delta U &= -T_{\rho}{}^{\rho(\mu}\lc{\nabla}^{\nu)}\phi\delta g_{\mu\nu} - 2\delta_{\mu}^{[\nu}\lc{\nabla}^{\rho]}\phi\delta\Gamma^{\mu}{}_{\nu\rho} + T_{\nu}{}^{\nu\mu}\lc{\nabla}_{\mu}\delta\phi\,,\\
\delta V &= -\lc{\nabla}^{(\mu}\phi Q^{\nu)\rho}{}_{\rho}\delta g_{\mu\nu} + g^{\mu\nu}\lc{\nabla}^{\rho}\phi\lc{\nabla}_{\rho}\delta g_{\mu\nu} - 2\delta_{\mu}^{\nu}\lc{\nabla}^{\rho}\phi\delta\Gamma^{\mu}{}_{\nu\rho} + Q^{\mu\nu}{}_{\nu}\lc{\nabla}_{\mu}\delta\phi\,,\\
\delta W &= \left(M^{\rho(\mu\nu)}\lc{\nabla}_{\rho}\phi + M^{\rho(\mu}{}_{\rho}\lc{\nabla}^{\nu)}\phi\right)\delta g_{\mu\nu} + g^{\rho(\mu}\lc{\nabla}^{\nu)}\phi\lc{\nabla}_{\rho}\delta g_{\mu\nu} - \left(\delta_{\mu}^{\rho}\lc{\nabla}^{\nu}\phi + g^{\nu\rho}\lc{\nabla}_{\mu}\phi\right)\delta\Gamma^{\mu}{}_{\nu\rho} + Q_{\nu}{}^{\nu\mu}\lc{\nabla}_{\mu}\delta\phi\,.
\end{align}
\end{subequations}
Collecting the different variation terms, we thus have
\begin{equation}
\delta S_{\text{ST}} = -\frac{1}{2\kappa^2}\int_M\left[A^{\mu\nu}\delta g_{\mu\nu} + B^{\rho\mu\nu}\lc{\nabla}_{\rho}\delta g_{\mu\nu} + C_{\mu}{}^{\nu\rho}\delta\Gamma^{\mu}{}_{\nu\rho} + D\delta\phi + E^{\mu}\lc{\nabla}_{\mu}\delta\phi\right]\sqrt{-g}\dd^4x\,,
\end{equation}
with the terms
\begin{subequations}
\begin{align}
A^{\mu\nu} &= \left(M^{\rho\sigma(\mu}M_{\sigma}{}^{\nu)}{}_{\rho} - M^{\rho(\mu\nu)}M^{\sigma}{}_{\rho\sigma}\right)L_G + \frac{1}{2}\lc{\nabla}^{\mu}\phi\lc{\nabla}^{\nu}\phi L_X - T_{\rho}{}^{\rho(\mu}\lc{\nabla}^{\nu)}\phi L_U - \lc{\nabla}^{(\mu}\phi Q^{\nu)\rho}{}_{\rho}L_V\nonumber\\
&\phantom{=}+ \left(M^{\rho(\mu\nu)}\lc{\nabla}_{\rho}\phi + M^{\rho(\mu}{}_{\rho}\lc{\nabla}^{\nu)}\phi\right)L_W + \frac{1}{2}g^{\mu\nu}L\,,\\
B^{\rho\mu\nu} &= \left(M^{\rho(\mu\nu)} - M^{\sigma(\mu}{}_{\sigma}g^{\nu)\rho} - M^{[\rho\sigma]}{}_{\sigma}g^{\mu\nu}\right)L_G + g^{\mu\nu}\lc{\nabla}^{\rho}\phi L_V + g^{\rho(\mu}\lc{\nabla}^{\nu)}\phi L_W\,,\\
C_{\mu}{}^{\nu\rho} &= \left(M^{\nu\sigma}{}_{\sigma}\delta_{\mu}^{\rho} + M^{\sigma}{}_{\mu\sigma}g^{\nu\rho} - M^{\nu\rho}{}_{\mu} - M^{\rho}{}_{\mu}{}^{\nu}\right)L_G - 2\delta_{\mu}^{[\nu}\lc{\nabla}^{\rho]}\phi L_U - 2\delta_{\mu}^{\nu}\lc{\nabla}^{\rho}\phi L_V - (\delta_{\mu}^{\rho}\lc{\nabla}^{\nu}\phi + g^{\nu\rho}\lc{\nabla}_{\mu}\phi)L_W\,,\\
D &= L_{\phi}\,,\\
E^{\mu} &= -\lc{\nabla}^{\mu}\phi L_X + T_{\nu}{}^{\nu\mu}L_U + Q^{\mu\nu}{}_{\nu}L_V + Q_{\nu}{}^{\nu\mu}L_W\,.
\end{align}
\end{subequations}
This gravitational part of the action further needs to be complemented by a matter action. For simplicity, we assume that the matter fields couple only to the metric, whereas there is no direct coupling to the teleparallel connection or the scalar field. It follows that the variation of the matter action with respect to the metric is of the general form
\begin{equation}\label{eq:metricmatactvar}
\delta S_{\text{m}} = \frac{1}{2}\int_M\Theta^{\mu\nu}\delta g_{\mu\nu}\sqrt{-g}\dd^4x\,,
\end{equation}
which defines the energy-momentum tensor \(\Theta_{\mu\nu}\). When deriving the field equations, we finally must also take into account the flatness condition~\eqref{eq:curvature}, from which follows that variations of the connection must be of the form~\cite{Hohmann:2021fpr}
\begin{equation}\label{eq:flatvar}
\delta\Gamma^{\mu}{}_{\nu\rho} = \nabla_{\rho}\xi^{\mu}{}_{\nu}\,.
\end{equation}
With this definition, we can perform integration by parts, after which we write the variation of the complete action in the form~\cite{Hohmann:2022mlc}
\begin{equation}
\delta S = \delta S_{\text{ST}} + \delta S_{\text{m}} = -\frac{1}{2\kappa^2}\int_M\left(\mathcal{G}^{\mu\nu}\delta g_{\mu\nu} + \mathcal{C}_{\mu}{}^{\nu}\xi^{\mu}{}_{\nu} + \mathcal{F}\delta\phi\right)\sqrt{-g}\dd^4x\,,
\end{equation}
where the newly introduced terms constitute the metric field equation
\begin{subequations}
\label{eq:generalfe}
\begin{equation}
0 = \mathcal{G}_{\mu\nu} = A_{\mu\nu} - \lc{\nabla}^{\rho}B_{\rho\mu\nu} - \kappa^2\Theta_{\mu\nu}\,,
\end{equation}
the connection field equation
\begin{equation}
0 = \mathcal{C}_{\mu}{}^{\nu} = M^{\sigma}{}_{\rho\sigma}C_{\mu}{}^{\nu\rho} - \nabla_{\rho}C_{\mu}{}^{\nu\rho}\,,
\end{equation}
as well as the scalar field equation
\begin{equation}
0 = \mathcal{F} = D - \lc{\nabla}_{\mu}E^{\mu}\,.
\end{equation}
\end{subequations}
With these field equations at hand, we can proceed with the post-Newtonian approximation, which we will perform in the following sections.

\section{Post-Newtonian approximation}\label{sec:ppn}
In this article, we make use of the parametrized post-Newtonian (PPN) formalism~\cite{Will:1993ns,Will:2014kxa,Will:2018bme}. This section provides a brief overview of the PPN formalism, keeping in mind that we aim to apply it to the generalized scalar-teleparallel theories of gravity discussed in the previous section. The PPN formalism relies on the assumption that the source of the gravitational field is a perfect fluid, which has a relatively small velocity measured in units of the speed of light in a specific fixed frame of reference. Additionally, it assumes that all the relevant physical quantities required to solve the gravitational field equations can be expanded in terms of this velocity. Here, we elaborate on how this expansion in velocity orders is carried out for the quantities needed in our calculations in the subsequent sections of this article.

The starting point of our calculation is the energy-momentum tensor of a perfect fluid with rest energy density \(\rho\), specific internal energy \(\Pi\), pressure \(p\) and four-velocity \(u^{\mu}\), which is given by
\begin{equation}\label{eq:tmunu}
\Theta^{\mu\nu} = (\rho + \rho\Pi + p)u^{\mu}u^{\nu} + pg^{\mu\nu}\,.
\end{equation}
The four-velocity \(u^{\mu}\) is normalized by the metric \(g_{\mu\nu}\), so that \(u^{\mu}u^{\nu}g_{\mu\nu} = -1\). We will now expand all dynamical quantities in orders \(\mathcal{O}(n) \propto |\vec{v}|^n\) of the velocity \(v^i = u^i/u^0\) of the source matter in a given frame of reference, starting with the field variables. The metric \(g_{\mu\nu}\) will be expanded around the flat Minkowski metric \(\eta_{\mu\nu}=\mathrm{diag}(-1, 1, 1, 1)\)
\begin{equation}\label{eq:metricexp}
g_{\mu\nu} = \eta_{\mu\nu} + h_{\mu\nu} = \eta_{\mu\nu} + \order{h}{1}_{\mu\nu} + \order{h}{2}_{\mu\nu} + \order{h}{3}_{\mu\nu} + \order{h}{4}_{\mu\nu} + \mathcal{O}(5)\,.
\end{equation}
For the teleparallel connection coefficients, we make an ansatz of the form
\begin{equation}
\Gamma^{\mu}{}_{\nu\rho} = (\Lambda^{-1})^{\mu}{}_{\sigma}\partial_{\rho}\Lambda^{\sigma}{}_{\nu}\,,
\end{equation}
and then expand
\begin{equation}\label{eq:connexp}
\Lambda^{\mu}{}_{\nu} = \delta^{\mu}_{\nu} + \lambda^{\mu}{}_{\nu} = \delta^{\mu}_{\nu} + \order{\lambda}{1}^{\mu}{}_{\nu} + \order{\lambda}{2}^{\mu}{}_{\nu} + \order{\lambda}{3}^{\mu}{}_{\nu} + \order{\lambda}{4}^{\mu}{}_{\nu} + \mathcal{O}(5)\,.
\end{equation}
Finally, we expand the scalar field \(\phi\)
\begin{equation}\label{eq:scalarexp}
\phi = \Phi + \psi = \Phi + \order{\psi}{1} + \order{\psi}{2} + \order{\psi}{3} + \order{\psi}{4} + \mathcal{O}(5)\,.
\end{equation}
around its cosmological background value \(\Phi\), which will be assumed to be constant. Here we have used overscript numbers to denote velocity orders, i.e., each term carrying an overscript \(n\) is of order \(\mathcal{O}(n)\). Velocity orders beyond the fourth order are not considered and will not be relevant for our calculation. Also up to the fourth velocity order not all components of the perturbations will be relevant. A detailed analysis, which involves expanding the equations of motion for test particles relevant for experimental tests of post-Newtonian gravity, the Newtonian energy-momentum conservation, dissipative effects such as the emission of gravitational radiation as well as the symmetry under time reversal in the absence of such dissipative effects at the lowest velocity orders, shows that not all components need to be expanded to the fourth velocity order, while others vanish; we omit these steps here for brevity, and refer the reader to a comprehensive discussion of this derivation~\cite{Will:1993ns}. It turns out that the only relevant, non-vanishing components of the field variables we need to determine in this article are given by
\begin{equation}\label{eq:ppnfields}
\order{h}{2}_{00}\,, \quad
\order{h}{2}_{ij}\,, \quad
\order{h}{3}_{i0}\,, \quad
\order{h}{4}_{00}\,, \quad
\order{\lambda}{2}_{00}\,, \quad
\order{\lambda}{2}_{ij}\,, \quad
\order{\lambda}{3}_{0i}\,, \quad
\order{\lambda}{3}_{i0}\,, \quad
\order{\lambda}{4}_{00}\,, \quad
\order{\psi}{2}\,, \quad
\order{\psi}{4}\,.
\end{equation}
Using the expansion~\eqref{eq:metricexp} of the metric tensor we can now also expand the energy-momentum tensor~\eqref{eq:tmunu} into velocity orders. For this purpose we must assign velocity orders also to the rest mass density, specific internal energy and pressure of the perfect fluid. Based on their orders of magnitude in the solar system one assigns velocity orders \(\mathcal{O}(2)\) to \(\rho\) and \(\Pi\) and \(\mathcal{O}(4)\) to \(p\). The energy-momentum tensor~\eqref{eq:tmunu} can then be expanded in the form
\begin{subequations}\label{eq:energymomentum}
\begin{align}
\Theta_{00} &= \rho\left(1 + \Pi + v^2 - 2\order{h}{2}_{00}\right) + \mathcal{O}(6)\,,\\
\Theta_{0j} &= -\rho v_j + \mathcal{O}(5)\,,\\
\Theta_{ij} &= \rho v_iv_j + p\delta_{ij} + \mathcal{O}(6)\,.
\end{align}
\end{subequations}
In the following section, we will derive the post-Newtonian expansion of the field equations, and solve them by increasing velocity orders.

\section{Solving the field equations}\label{sec:solution}
In order to obtain the post-Newtonian perturbations of the dynamical fields for a given matter source, we now perform an expansion of the gravitational field equations up to the required perturbation order. For this purpose, we need to perform a Taylor expansion of the free function \(L\) in the gravitational action~\eqref{eq:staction} around the background values of the dynamical fields, which we discussed in the previous section. In the following, quantities of the form \(l, l_i, l_{ij}, l_{ijk} \ldots\) will denote the value of \(L, L_i, L_{ij}, L_{ijk}, \ldots\) evaluated at the background \(\phi = \Phi, g_{\mu\nu} = \eta_{\mu\nu}, \Gamma^{\mu}{}_{\nu\rho} = 0\), and hence \(G = X = U = V = W = 0\), where \(i, j, k \in \{G, X, U, V, W, \phi\}\). We will then solve the field equations by making an ansatz for the perturbations as a linear combination of the so-called post-Newtonian potentials with unknown constant coefficients, which are then determined by solving the field equations. Our derivation progresses order by order in the post-Newtonian expansion. We start with the zeroth order, which corresponds to the background solution, shown in section~\ref{ssec:order0}. Subsequently, section~\ref{ssec:order2} presents the solution for the second order, followed by section~\ref{ssec:order3} for the third order, and finally section~\ref{ssec:order4} for the fourth order.

\subsection{Zeroth velocity order}\label{ssec:order0}
We start our discussion with the zeroth order of the field equations~\eqref{eq:generalfe}. From the expansion~\eqref{eq:energymomentum} follows that at the zeroth velocity order the energy-momentum tensor vanishes, \(\order{\Theta}{0}_{\mu\nu} = 0\), so that we are left with solving the vacuum field equations. Inserting our assumed background values $g_{\mu\nu}=\eta_{\mu\nu}$ and $\phi=\Phi$ into the respective field equations, we find that they take the form
\begin{equation}\label{eq:order0}
0 = -\frac{1}{2}l\eta_{\mu\nu},\qquad 0=l_\phi\,.
\end{equation}
It thus follows that the field equations are solved at the zeroth order only for theories which satisfy \(l =l_\phi= 0\).  To maintain simplicity, we assume a massless scalar field to avoid solutions expressed in terms of Yukawa-type potentials. This simplification can be achieved by setting both \(l_{\phi\phi}= 0\) and \(l_{\phi\phi\phi}= 0\) equal to zero. Henceforth, throughout the subsequent sections of this article, these assumptions will be employed.

\subsection{Second velocity order}\label{ssec:order2}
We now continue with the second order metric, scalar field and connection equations. We find that the only non-vanishing components are expressed as
\begin{subequations}\label{eq:2order}
\begin{align}
\order{\mathcal{G}}{2}_{00} &= -\kappa^2\rho + l_V\triangle\order{\psi}{2} + \frac{1}{2}l_G\left(\order{h}{2}^{ab}{}_{,ab} - \triangle\order{h}{2}^a{}_a\right)\,,\label{eq:e002}\\
\order{\mathcal{G}}{2}_{ab} &= -l_W\order{\psi}{2}_{,ab} - l_V\delta_{ab}\triangle\order{\psi}{2} + \frac{1}{2}l_G\left(\order{h}{2}_{00,ab} - \order{h}{2}^{c}{}_{c,ab} + \order{h}{2}^c{}_{b,ac} +\order{h}{2}^c{}_{a,bc} - \triangle\order{h}{2}_{ab} - \delta_{ab}\left(\triangle\order{h}{2}_{00} + \order{h}{2}^{cd}{}_{,cd} - \triangle\order{h}{2}^c{}_c\right)\right)\,,\label{eq:eij2}\\
\order{\mathcal{C}}{2}_0{}^0 &= (l_U + 2l_V)\triangle\order{\psi}{2}\,,\label{eq:n002}\\
\order{\mathcal{C}}{2}_a{}^b &= (2l_W - l_U)\order{\psi}{2}_{,a}{}^b + (l_U + 2l_V)\delta_a^b\triangle\order{\psi}{2}\,,\label{eq:nab2}\\
\order{\mathcal{F}}{2} &= l_X\triangle\order{\psi}{2} + l_W\left(2\order{\lambda}{2}^{ab}{}_{,ab} - \order{h}{2}^{ab}{}_{,ab}\right) + l_V\triangle\left(\order{h}{2}_{00} - \order{h}{2}^a{}_a + 2\order{\lambda}{2}^0{}_0 + 2\order{\lambda}{2}^a{}_a\right) + l_U\left(\triangle\order{\lambda}{2}^0{}_0 + \triangle\order{\lambda}{2}^a{}_a - \order{\lambda}{2}^{ab}{}_{,ab}\right)\,,\label{eq:s2}
\end{align}
\end{subequations}
where \(\triangle = \eta^{ab}\partial_a\partial_b = \delta^{ab}\partial_a\partial_b\) is the spatial Laplace operator of the flat background metric~\cite{Will:1993ns}. These equations can be solved with the ansatz
\begin{equation}\label{eq:ansatz2}
\order{h}{2}_{00} = a_1U\,, \quad
\order{h}{2}_{ab} = a_2U\delta_{ab} + a_3U_{ab}\,, \quad
\order{\lambda}{2}^0{}_0 = a_4U\,, \quad
\order{\lambda}{2}^a{}_b = a_5\delta_b^aU+a_6U^a{}_b\,, \quad
\order{\psi}{2}= a_7U\,.
\end{equation}
Here $a_i$ are constant coefficients, which we will determine by solving the field equations, while $U$ and $U_{ab}$ are post-Newtonian functionals of the matter variables. These functionals are related to the matter variables by the differential relation
\begin{equation}\label{eq:functionals2}
\triangle\chi = -2U\,, \quad
U_{ab} = \chi_{,ab} + U\delta_{ab}\,, \quad
\triangle U = -4\pi\rho\,,
\end{equation}
where \(\chi\) is the so-called superpotential, which is auxiliary in the definition of \(U_{ab}\). Inserting the ansatz~\eqref{eq:ansatz2} into the second order field equations~\eqref{eq:2order}, one obtains a linear system of equations for the constant coefficients, which is given by
\begin{subequations}\label{eq:cond2}
\begin{align}
l_G(a_2 + a_3) - l_Va_7 &= \frac{\kappa^2}{4\pi}\,,\\
l_G(a_1 - a_2 - a_3) - 2l_Wa_7 &= 0\,,\\
l_G(a_1 - a_2 - a_3) + 2l_Va_7 &= 0\,,\\
(l_U + 2l_V)a_7 &= 0\,,\\
(l_U - 2l_W)a_7 &= 0\,,\\
l_W(a_2 - a_3 - 2a_5 + 2a_6) - l_V(a_1 - 3a_2 - a_3 + 2a_4 + 6a_5 + 2a_6) - l_U(a_4 + 2a_5 + 2a_6) - l_Xa_7 &= 0\,.
\end{align}
\end{subequations}
In order to solve these equations, we need to supplement them with another equation, which fixes the gauge freedom left by infinitesimal coordinate transformations, and which allows us to set
\begin{equation}
a_3 = 0\,.
\end{equation}
In order to determine the possible solutions to the resulting system of linear equations for the coefficients \(a_1, \ldots, a_7\), we then have to distinguish a number of different cases. For this purpose, it is most convenient to simplify the system first by taking suitable linear combinations, which are chosen carefully such that the obtained system remains equivalent to the original system regardless of the values of the constant Taylor coefficients. During this procedure, one finds that one equation is redundant and can thus be eliminated. The resulting system can then be written in matrix form as
\begin{equation}\label{eq:matcond2}
\mat{M} \cdot \vek{A} = \vek{T}\,,
\end{equation}
where we introduced the matrix
\begin{equation}
\mat{M} = \begin{pmatrix}
l_G & 0 & 0 & 0 & 0 & 0 & l_V\\
0 & l_G & 0 & 0 & 0 & 0 & -l_V\\
0 & 0 & 1 & 0 & 0 & 0 & 0\\
0 & 0 & 0 & 0 & 0 & 0 & l_U + 2l_V\\
0 & 0 & 0 & 0 & 0 & 0 & l_V + l_W\\
l_V & -3l_V - l_W & 0 & l_U + 2l_V & 2l_U + 6l_V + 2l_W & 2l_U + 2l_V - 2l_W & l_X
\end{pmatrix}
\end{equation}
and defined the vectors
\begin{equation}
\vek{A}^t = \begin{pmatrix}
a_1 &
a_2 &
a_3 &
a_4 &
a_5 &
a_6 &
a_7
\end{pmatrix}\,, \quad
\vek{T}^t = \frac{\kappa^2}{4\pi}\begin{pmatrix}
1 &
1 &
0 &
0 &
0 &
0
\end{pmatrix}\,.
\end{equation}
This system possesses at least one solution if and only if \(\vek{T}\) lies in the image of \(\mat{M}\), and the dimension of the solution space is that of the kernel of \(\mat{M}\), which is related to its rank. In the generic case, where we do not assume any particular relation between the Taylor coefficients determining the perturbed field equations, \(\mat{M}\) has rank 5 and yields the solution
\begin{equation}\label{eq:coeff2r5}
a_1 = a_2 = \frac{\kappa^2}{4\pi l_G}, \quad
a_3 = a_7 = 0\,,
\end{equation}
together with the remaining equation
\begin{equation}\label{eq:gen2}
(l_U + 2l_V)a_4 + 2(l_U + 3l_V + l_W)a_5 + 2(l_U + l_V - l_W)a_6 = \frac{\kappa^2(2l_V + l_W)}{4\pi l_G}\,,
\end{equation}
which can be solved for any of the remaining variables, provided that the corresponding coefficient is non-vanishing. Note that this case is obtained if and only if \(l_G \neq 0\) and at least one of the linear combinations \(l_U + 2l_V\) or \(l_V + l_W\) is non-vanishing. We see that this condition is equivalent to demanding that at least one of the terms in the equation~\eqref{eq:gen2} is non-vanishing.

We can then come to cases where the parameters take particular values. First note that also in these cases \(l_G\) must be non-vanishing, since otherwise the first two equations would contradict each other, and so the system has no solutions for \(l_G = 0\). We will therefore demand \(l_G \neq 0\) for the remainder of this article. If
\begin{equation}
l_U + 2l_V = l_V + l_W = 0\,,
\end{equation}
we find that the coefficients \(a_4, a_5, a_6\) do not contribute to the equations. The rank of \(\mat{M}\) reduces to 4 and the remaining parameters are determined by the unique solution
\begin{equation}\label{eq:coeff2r4}
a_1 = \frac{\kappa^2}{\pi l_G}\frac{l_U^2 - l_Gl_X}{3l_U^2 - 4l_Gl_X}\,, \quad
a_2 = \frac{\kappa^2}{2\pi l_G}\frac{l_U^2 - 2l_Gl_X}{3l_U^2 - 4l_Gl_X}\,, \quad
a_3 = 0\,, \quad
a_7 = \frac{\kappa^2}{2\pi}\frac{l_U}{3l_U^2 - 4l_Gl_X}\,.
\end{equation}
Note that this solution exists only if the appearing denominator is non-vanishing. If it vanishes, the rank of the matrix \(\mat{M}\) further reduces to 3, since the last row of \(\mat{M}\) becomes a linear combination of the first two. One finds that the corresponding linear equations in general contradict each other, unless \(l_X = l_U = 0\), for which \(a_7\) disappears from the equations, and one has the unique solution
\begin{equation}\label{eq:coeff2r3}
a_1 = a_2 = \frac{\kappa^2}{4\pi l_G}\,, \quad
a_3 = 0\,.
\end{equation}
Any further restriction of the parameter values would again lead to an inconsistent system of equations, and so we find that these are the solutions in all possible cases. In summary, we find that between two and four coefficients in the solution remain undetermined from the field equations, depending on the values of the Taylor coefficients. We will discuss the relevance of these undetermined parameters when we come to the full solution at higher velocity orders and the resulting post-Newtonian metric.

\subsection{Third velocity order}\label{ssec:order3}
At the third velocity order the only non-vanishing components of the field equations are
\begin{subequations}\label{eq:3order}
\begin{align}
\order{\mathcal{G}}{3}_{0a} &= \kappa^2\rho v_a - l_W\order{\psi}{2}_{,0a} - \frac{1}{2}l_G\left(\order{h}{2}^b{}_{b,0a} + \triangle\order{h}{3}_{0a} - \order{h}{3}^b{}_{0,ab} - \order{h}{2}^b{}_{a,0b}\right)\,,\\
\order{\mathcal{C}}{3}_0{}^b &= (2l_W - l_U)\order{\psi}{2}_{,0}{}^b\,,\\
\order{\mathcal{C}}{3}_a{}^0 &= (l_U - 2l_W)\order{\psi}{2}_{,0a}\,.
\end{align}
\end{subequations}
To solve these equations we use the ansatz
\begin{equation}\label{eq:ansatz3}
\order{h}{3}_{0a} = a_8V_a + a_9W_a\,, \quad
\order{\lambda}{3}^0{}_a = a_{10}V_a + a_{11}W_a\,, \quad
\order{\lambda}{3}^a{}_0 = a_{12}V^a + a_{13}W^a\, ,
\end{equation}
with the potentials satisfying
\begin{equation}
\triangle V_a = -4\pi\rho v_a\,, \quad
\triangle W_a = -4\pi\rho v_a + 2U_{,0a}\,.
\end{equation}
Note that \(\order{\lambda}{3}^0{}_a\) and \(\order{\lambda}{3}^a{}_0\) do not appear in the third order field equations; nevertheless, we introduce an ansatz here with unknown coefficients, since they may appear in higher order field equations, which would then allow us to determine these coefficients as well. With this ansatz and the solution of the second order field equations shown in the previous section, we thus obtain only a single equation among the coefficients which is given by
\begin{equation}\label{eq:coeff3}
a_8 + a_9 = -\frac{\kappa^2}{2\pi l_G}\,,
\end{equation}
independently of the values of the parameter values discussed in the previous section. Note that the connection field equations are always solved identically, since for theories with \(l_U - 2l_W \neq 0\), we have obtained \(\order{\psi}{2} = 0\) at the second velocity order. It follows that the equations depend only on the Taylor coefficient \(l_G\), which we have demanded to be non-vanishing while solving the second order equations, which otherwise would not possess any solution. We see that this equation is always insufficient in order to solve for all coefficients, which is again a consequence of the gauge freedom resulting from infinitesimal coordinate transformations. Here we will not make a gauge choice, and defer this step to the fourth order, as it is conventional to define the standard PPN gauge by the non-appearance of a particular term in the fourth order metric solution. We can thus only determine the linear combination \(a_8 + a_9\). Hence, we are left with five undetermined parameters in the solution, which must be taken into account when we come to the fourth velocity order.

\subsection{Fourth velocity order}\label{ssec:order4}
At the fourth order, the equations to be solved are given by $\order{\mathcal{G}}{4}_{00}$, $\order{\mathcal{G}}{4}_{ab}$, $\order{\mathcal{F}}{4}$, $\order{\mathcal{C}}{4}_0{}^0$, $\order{\mathcal{C}}{4}_a{}^b$, similarly to the second velocity order. We will not show these equations here, as they turn out to be rather lengthy. These equations contain second order derivatives of the fourth order perturbations $\order{h}{4}_{00}$, $\order{h}{4}_{ab}$, $\order{\mathcal{\psi}}{4}$, $\order{\lambda}{4}^0{}_0$, $\order{\lambda}{4}^a{}_b$. Note that only the first of these needs to be determined in the PPN formalism, in order to obtain the PPN parameters. Instead of making an ansatz for all of the aforementioned perturbations, as we did for the second velocity order, we will therefore use a different method, and eliminate all irrelevant terms from the field equations, by considering a suitable linear combination. In order to achieve this, we will use the equations
\begin{equation}
\triangle\order{\mathcal{G}}{4}_{00}\,, \quad
\triangle\order{\mathcal{G}}{4}_a{}^a\,, \quad
\partial_a\partial_b\order{\mathcal{G}}{4}^{ab}\,, \quad
\triangle\order{\mathcal{C}}{4}_0{}^0\,, \quad
\triangle\order{\mathcal{C}}{4}_a{}^a\,, \quad
\partial_a\partial_b\order{\mathcal{C}}{4}^{ab}\,, \quad
\triangle\order{\mathcal{F}}{4}\,,
\end{equation}
which are now scalars, and which contain the scalar quantities
\begin{equation}
\triangle\triangle\order{h}{4}_{00}\,, \quad
\triangle\triangle\order{h}{4}_a{}^a\,, \quad
\triangle\partial_a\partial_b\order{h}{4}^{ab}\,, \quad
\triangle\triangle\order{\lambda}{4}^0{}_0\,, \quad
\triangle\triangle\order{\lambda}{4}^a{}_a\,, \quad
\triangle\partial_a\partial_b\order{\lambda}{4}^{ab}\,, \quad
\triangle\triangle\order{\psi}{4}\,.
\end{equation}
In order to eliminate all of these terms except for the first one, we need to take a suitable linear combination of the aforementioned equations. To find this linear combination, it is most convenient to write the general form of the equations in matrix form as
\begin{equation}
\begin{pmatrix}
\triangle\order{\mathcal{G}}{4}_{00}\\
\triangle\order{\mathcal{G}}{4}_a{}^a\\
\partial_a\partial_b\order{\mathcal{G}}{4}^{ab}\\
\triangle\order{\mathcal{C}}{4}_0{}^0\\
\triangle\order{\mathcal{C}}{4}_a{}^a\\
\partial_a\partial_b\order{\mathcal{C}}{4}^{ab}\\
\triangle\order{\mathcal{F}}{4}
\end{pmatrix} = \begin{pmatrix}
0 & -\frac{l_G}{2} & \frac{l_G}{2} & 0 & 0 & 0 & l_V\\
-l_G & \frac{l_G}{2} & -\frac{l_G}{2} & 0 & 0 & 0 & -3l_V - l_W\\
0 & 0 & 0 & 0 & 0 & 0 & -l_V - l_W\\
0 & 0 & 0 & 0 & 0 & 0 & l_U + 2l_V\\
0 & 0 & 0 & 0 & 0 & 0 & 2l_U + 6l_V + 2l_W\\
0 & 0 & 0 & 0 & 0 & 0 & 2l_V + 2l_W\\
l_V & -l_V & -l_W & l_U + 2l_V & l_U + 2l_V & -l_U + 2l_W & l_X
\end{pmatrix} \cdot \begin{pmatrix}
\triangle\triangle\order{h}{4}_{00}\\
\triangle\triangle\order{h}{4}_a{}^a\\
\triangle\partial_a\partial_b\order{h}{4}^{ab}\\
\triangle\triangle\order{\lambda}{4}^0{}_0\\
\triangle\triangle\order{\lambda}{4}^a{}_a\\
\triangle\partial_a\partial_b\order{\lambda}{4}^{ab}\\
\triangle\triangle\order{\psi}{4}
\end{pmatrix} + \vek{\order{\mathcal{S}}{4}}\,,
\end{equation}
where \(\vek{\order{\mathcal{S}}{4}}\) contains all source terms, e.g., terms which originate from lower order perturbations, such as products of two second order perturbations or time derivatives of third order perturbations, as well as the fourth order matter terms. Once a Poisson-like equation for \(\order{h}{4}_{00}\) is obtained, it can be solved using the standard PPN ansatz
\begin{equation}\label{eq:ansatz4}
\order{h}{4}_{00} = a_{14}U^2 + a_{15}\Phi_1 + a_{16}\Phi_2 + a_{17}\Phi_3 + a_{18}\Phi_4 + a_{19}\Phi_W + a_{20}\mathcal{A} + a_{21}\mathcal{B}\,,
\end{equation}
which contains the fourth order scalar potentials
\begin{equation}
\triangle\Phi_1 = -4\pi\rho v^2\,, \quad
\triangle\Phi_2 = -4\pi\rho U\,, \quad
\triangle\Phi_3 = -4\pi\rho\Pi\,, \quad
\triangle\Phi_4 = -4\pi p\,,
\end{equation}
and \(\Phi_W, \mathcal{A}, \mathcal{B}\) are defined in~\cite{Will:1993ns}. In order to obtain the post-Newtonian limit, we need to determine the values of the constant coefficients \(a_{14}, \ldots, a_{21}\), in addition to any yet undetermined coefficients appearing in the lower order metric perturbations. As in the lower orders, also here we have a remaining gauge freedom at the fourth order, which is used to set the standard PPN gauge \(a_{21} = 0\). The final solution, as well as the aforementioned steps used to solve the fourth order equations, depend on the values of the parameter functions, as it will be different for the individual branches introduced in section~\ref{ssec:order2}. Since this procedure is rather non-trivial, we divide it into separate subsections.

\subsubsection{Rank 5}
We first study the case that at least one of \(l_U + 2l_V\) or \(l_U - 2l_W\) is non-vanishing, so that the matrix \(\mat{M}\) in section~\ref{ssec:order2} is of rank 5. In this case one finds that after substituting the second and third order solutions, the four equations \(\partial_a\partial_b\order{\mathcal{G}}{4}^{ab}\), \(\triangle\order{\mathcal{C}}{4}_0{}^0\), \(\triangle\order{\mathcal{C}}{4}_a{}^a\), \(\partial_a\partial_b\order{\mathcal{C}}{4}^{ab}\) (including the lower order terms not shown here) are linearly dependent, so that there exists only one non-trivial equation which contains no other fourth order field variables besides \(\triangle\triangle\order{\psi}{4}\). This equation can thus be used to eliminate the fourth order scalar field from the linear combination \(\triangle\order{\mathcal{G}}{4}_{00} + \triangle\order{\mathcal{G}}{4}_a{}^a\), in which only \(\triangle\triangle\order{h}{4}_{00}\) and \(\triangle\triangle\order{\psi}{4}\) appear. The resulting equation can then be solved for \(\triangle\triangle\order{h}{4}_{00}\), which together with the gauge condition \(a_{21} = 0\) yields the solution
\begin{equation}\label{eq:coeff4gr}
-\frac{4a_{8}}{7} = -4a_9 = \frac{a_{15}}{2} = a_{17} = \frac{a_{18}}{3} = \frac{\kappa^2}{4\pi l_G}\,, \quad
-2a_{14} = a_{16} = \frac{\kappa^4}{16\pi^2l^2_G}\,, \quad
a_{19} = a_{20} = a_{21} = 0\,,
\end{equation}
so that now all metric components up to the fourth velocity order are determined.

\subsubsection{Rank 4}
In the case
\begin{equation}
l_U + 2l_V = l_V + l_W = 0\,, \quad
3l_U^2 - 4l_Gl_X \neq 0\,,
\end{equation}
in which the matrix \(\mat{M}\) in section~\ref{ssec:order2} has rank 4, we find that the three equations \(\triangle\order{\mathcal{G}}{4}_{00}\), \(\triangle\order{\mathcal{G}}{4}_a{}^a\), \(\triangle\order{\mathcal{F}}{4}\) form a non-degenerate linear system for the three terms
\begin{equation}
\triangle\triangle\order{h}{4}_{00}\,, \quad
\triangle\triangle\order{h}{4}_a{}^a - \triangle\partial_a\partial_b\order{h}{4}^{ab}\,, \quad
\triangle\triangle\order{\psi}{4}\,,
\end{equation}
and can therefore be solved for the first term. The remaining four equations do not contain any fourth order potentials. However, in contrast to the rank 5 case discussed above, they are not solved identically after substituting the lower order solution, since in this case the lower order field equations have left further coefficients \(a_4, a_5, a_6\) in the lower order perturbations undetermined. We therefore now obtain further conditions on these yet undetermined lower order coefficients, as well as consistency conditions on the Taylor coefficients of the Lagrangian. These conditions read
\begin{equation}
l_U^2(l_{\phi U} + 2l_{\phi V}) = l_U^2(l_{\phi V} + l_{\phi W}) = l_U(l_U - 2l_{\phi G})a_6 = 0\,.
\end{equation}
Note that the last condition can always be solved by \(a_6 = 0\), which will have no further influence on the solution of the remaining equations, and thus does not pose any restriction on the Taylor coefficients. The remaining equations can either be solved by \(l_U = 0\), which will lead to the same solution~\eqref{eq:coeff4gr} as in the previous case, or by
\begin{equation}
l_{\phi U} + 2l_{\phi V} = l_{\phi V} + l_{\phi W} = 0\,,
\end{equation}
for which, together with the gauge condition \(a_{21} = 0\), one finds the solution
\begin{subequations}\label{eq:coeff4nt}
\begin{align}
a_8 &= -\frac{\kappa^2}{4\pi l_G}\frac{5l_U^2 - 7l_Gl_X}{3l_U^2 - 4l_Gl_X}\,,\\
a_9 &= -\frac{\kappa^2}{4\pi l_G}\frac{l_U^2 - l_Gl_X}{3l_U^2 - 4l_Gl_X}\,,\\
a_{14} &= -\frac{\kappa^4}{16\pi^2l_G^2}\frac{21l_U^6 - 73l_U^4l_Gl_X + 4l_U^2l_Gl_X^2(l_{\phi U} + 21l_X) - 32l_X^3l_G^3 + 6l_U^5l_{\phi G} + 8l_Ul_X^2l_G^2l_{\phi G} - 2l_U^3l_G(l_Gl_{\phi X} + 8l_Xl_{\phi G})}{(3l_U^2 - 4l_Gl_X)^3}\,,\\
a_{15} &= \frac{\kappa^2}{2\pi l_G}\,,\\
a_{16} &= \frac{\kappa^4}{8\pi^2l_G^2}\frac{9l_U^6 - 45l_U^4l_Gl_X - 4l_U^2l_Gl_X^2(l_{\phi U} - 17l_X) - 32l_X^3l_G^3 + 6l_U^5l_{\phi G} + 8l_Ul_X^2l_G^2l_{\phi G} + 2l_U^3l_G(l_Gl_{\phi X} + 8l_Xl_{\phi G})}{(3l_U^2 - 4l_Gl_X)^3}\,,\\
a_{17} &= \frac{\kappa^2}{\pi l_G}\frac{l_U^2 - l_Gl_X}{3l_U^2 - 4l_Gl_X}\,,\\
a_{18} &= \frac{3\kappa^2}{2\pi l_G}\frac{l_U^2 - 2l_Gl_X}{3l_U^2 - 4l_Gl_X}\,,\\
a_{19} &= a_{20} = a_{21} = 0\,,
\end{align}
\end{subequations}
which then again determines all metric components up to the fourth velocity order.

\subsubsection{Rank 3}
We finally come to the case
\begin{equation}
l_U = l_V = l_W = l_X = 0\,,
\end{equation}
in which the matrix \(\mat{M}\) in section~\ref{ssec:order2} has rank 3. In this case one finds that the linear combination
\begin{equation}
\triangle\order{\mathcal{G}}{4}_{00} + \triangle\order{\mathcal{G}}{4}_a{}^a
\end{equation}
can be solved for \(\order{h}{4}_{00}\), while the remaining fourth order equations do not contain any fourth order perturbations. As in the previously discussed rank 4 case, these equations, unless identically vanishing and thus trivially solved, thus pose an additional condition on the lower order perturbations, which have not been fully determined by the corresponding lower order field equations. Evaluating the independent terms in these equations, one finds the conditions
\begin{equation}
l_{\phi G}a_6 = (l_{\phi U} + 2l_{\phi V})a_7 = (l_{\phi V} + l_{\phi W})a_7 = l_{\phi X}a_7^2 + \frac{\kappa^4l_{\phi G}}{16\pi^2l_G^2} = \left(l_{\phi X}a_7 + \frac{\kappa^2l_{\phi U}}{8\pi l_G}\right)a_7 = 0\,.
\end{equation}
Depending on the values of the Taylor coefficients, this system of equations possesses different solutions. If \(l_{\phi G} \neq 0\), then we must have \(a_6 = 0\) and
\begin{equation}
a_7 = \pm\frac{\kappa^2}{4\pi l_G}\sqrt{-\frac{l_{\phi G}}{l_{\phi X}}}\,,
\end{equation}
which necessitates that \(l_{\phi X} \neq 0\) such that \(l_{\phi X}\) and \(l_{\phi G}\) have opposite signs, while further
\begin{equation}
l_{\phi U} = -2l_{\phi V} = 2l_{\phi W} = -8\pi l_Gl_{\phi X}a_7\,.
\end{equation}
Otherwise, if \(l_{\phi G} = 0\), we find that \(a_6\) remains undetermined. If
\begin{equation}\label{eq:lpuvwx}
l_{\phi U} = l_{\phi V} = l_{\phi W} = l_{\phi X} = 0\,,
\end{equation}
then also \(a_7\) is undetermined. Finally, if any of the coefficients~\eqref{eq:lpuvwx} is non-vanishing, it imposes \(a_7 = 0\). One finds that in all of the aforementioned cases the solution for the perturbation coefficients reduces to the solution~\eqref{eq:coeff4gr} found in the rank 5 case.

\section{PPN metric and parameters}\label{sec:ppnpar}
Using the solution for the metric perturbations obtained in the previous section, we can now determine the PPN parameters. For this purpose, we compare the metric perturbations~\eqref{eq:ansatz2}, \eqref{eq:ansatz3} and~\eqref{eq:ansatz4} to the standard PPN form of the metric, which is given by~\cite{Will:1993ns,Will:2014kxa}
\begin{subequations}\label{eq:ppnmetric}
\begin{align}
\order{h}{2}_{00} &= 2U\,,\\
\order{h}{2}_{ij} &= 2\gamma U\delta_{ij}\,,\\
\order{h}{3}_{0i} &= -\frac{1}{2}(3 + 4\gamma + \alpha_1 - \alpha_2 + \zeta_1 - 2 \xi )V_i - \frac{1}{2}(1 + \alpha_2 - \zeta_1 + 2\xi)W_i\,,\\
\order{h}{4}_{00} &= -2\beta U^2 - 2\xi \Phi_W + (2 + 2\gamma + \alpha_3 + \zeta_1  -2\xi)\Phi_1 + 2(1 + 3\gamma - 2\beta + \zeta_2 + \xi)\Phi_2\nonumber\\
&\phantom{=}+ 2(1 + \zeta_3)\Phi_3 + 2(3\gamma + 3\zeta_4 - 2\xi)\Phi_4 - (\zeta_1 - 2\xi) \mathcal{A}\,,
\end{align}
\end{subequations}
It is evident that this form of the metric is already in the PPN gauge, as \(\mathcal{B}\) is absent from the component \(\order{h}{4}_{00}\). Note in particular that the form of the component \(\order{h}{2}_{00}\) imposes the normalization condition \(a_1 \equiv 2\) for the effective gravitational constant. This can be understood as a choice of units, which is implemented by solving the normalization condition for \(\kappa\), and substituting the result in the remaining coefficients in the metric. Equivalently, one can absorb the normalization into a rescaling of the energy-momentum tensor by \(2/a_1\). This means that the coefficients of the terms \(U^2\) and \(\Phi_2\), which are quadratic in the matter source, incur a factor \(4/a_1^2\), while all other terms, which are linear in the matter source, incur a factor \(2/a_1\). Solving for the PPN parameters then yields the explicit solution
\begin{subequations}
\begin{align}
\beta &= -2\frac{a_{14}}{a_1^2}\,,\\
\gamma &= \frac{a_2}{a_1}\,,\\
\xi &= -\frac{a_{19}}{a_1}\,,\\
\alpha_1 &= -4\left(\frac{a_2 + a_8 + a_9}{a_1} + 1\right)\\
\alpha_2 &= -\frac{a_1 + 4a_9 + 2a_{20}}{a_1}\,,\\
\alpha_3 &= -\frac{2a_1 + a_2 - 2a_{15} - 2a_{20}}{a_1}\,,\\
\zeta_1 &= -2\frac{a_{19} + a_{20}}{a_1}\,,\\
\zeta_2 &= \frac{2a_{16} - 4a_{14}}{a_1^2} + \frac{a_{19} - 3a_2}{a_1} - 1\,,\\
\zeta_3 &= \frac{a_{17}}{a_1} - 1\,,\\
\zeta_4 &= -\frac{3a_2 - a_{18} + 2a_{19}}{3a_1}\,.
\end{align}
\end{subequations}
With these formulas at hand, we can now determine the PPN parameters from the metric solutions we have found before. We first note that for all cases which admit a unique metric solution up to the fourth velocity order, we find the values
\begin{equation}
\xi = \alpha_1 = \alpha_2 = \alpha_3 = \zeta_1 = \zeta_2 = \zeta_3 =\zeta_4 = 0\,,
\end{equation}
from which we deduce that there is no violation of the conservation of total energy-momentum, as well as no preferred frame or preferred location effects; theories of this type are called fully conservative. For the two parameters \(\beta\) and \(\gamma\), we find that we must distinguish two different cases. In the rank 5 and rank 3 cases, in which the coefficients in the metric perturbations take the form~\eqref{eq:coeff2r5} or~\eqref{eq:coeff2r3}, together with~\eqref{eq:coeff4gr}, we find the normalization condition
\begin{equation}\label{eq:newtongr}
\kappa^2 = 8\pi l_G\,,
\end{equation}
as well as the PPN parameters
\begin{equation}\label{eq:ppnpargr}
\beta = \gamma = 1\,,
\end{equation}
as in GR. The only potentially deviating result is obtained in the rank 4 case with the coefficients given by~\eqref{eq:coeff2r4} and~\eqref{eq:coeff4nt}, so that one finds the values
\begin{equation}\label{eq:newtonnt}
\kappa^2 = 2\pi l_G\frac{3l_U^2 - 4l_Gl_X}{l_U^2 - l_Gl_X}\,,
\end{equation}
and the PPN parameters are given by
\begin{subequations}\label{eq:ppnparnt}
\begin{align}
\beta - 1 &= \frac{l_U[7l_U^3l_Gl_X - 3l_U^5 - 2l_Ul_G^2(l_Ul_{\phi X} + 2l_X^2 - 2l_Xl_{\phi U}) + 2l_{\phi G}(l_U^2 - 2l_Gl_X)(3l_U^2 - 2l_Gl_X)]}{8(3l_U^2 - 4l_Gl_X)(l_U^2 - l_Gl_X)^2}\,,\\
\gamma - 1 &= -\frac{l_U^2}{2(l_U^2 - l_Gl_X)}\,.
\end{align}
\end{subequations}
Note that in the former case, the theory is indistinguishable from GR by its PPN parameters, and other experimental tests must be employed in order to distinguish them by observations. In the latter case, one finds deviating values for two parameters. In order to be consistent with measurements of these parameters, their deviation from the GR values \(\beta = \gamma = 1\) must be at most of the order of magnitude \(10^{-5}\)~\cite{Verma:2013ata}, and so we find that the corresponding combinations~\eqref{eq:ppnparnt} of the Taylor coefficients are highly constrained by observations.

\section{Classification of theories}\label{sec:classes}
\begin{figure}[!htbp]
\tikzsetfigurename{classes}
\begin{tikzpicture}[decision/.style={draw,rectangle,fill=black!10!white,outer sep=2pt},result/.style={draw,rectangle,rounded corners,outer sep=2pt},every edge/.style={draw,-Stealth},good/.style={fill=green!25!white},maybe/.style={fill=yellow!25!white},noppn/.style={fill=red!25!white},degen/.style={fill=magenta!25!white}]
\node[decision] (l0lp) at (0,0) {$l_0, l_{\phi}$};
\node[decision] (lpplppp) at (0,-1.5) {$l_{\phi\phi}, l_{\phi\phi\phi}$};
\node[decision] (lg) at (0,-3) {$l_G$};
\node[decision] (lulv) at (0,-4.5) {$l_U + 2l_V$};
\node[decision] (lvlw) at (0,-6) {$l_V + l_W$};
\node[decision] (lulglx) at (0,-7.5) {$3l_U^2 - 4l_Gl_X$};
\node[decision] (lu) at (3,-7.5) {$l_U$};
\node[decision] (lpulpv) at (3,-9) {$l_{\phi U} + 2l_{\phi V}$};
\node[decision] (lpvlpw) at (3,-10.5) {$l_{\phi V} + l_{\phi W}$};
\node[decision] (lx) at (0,-9) {$l_X$};
\node[decision] (lpg) at (0,-10.5) {$l_{\phi G}$};
\node[decision] (lpx) at (0,-12) {$l_{\phi X}$};
\node[decision] (lpglpx) at (0,-13.5) {$l_{\phi G}l_{\phi X}$};
\node[decision] (lpulpv2) at (0,-15) {$l_{\phi U} + 2l_{\phi V}$};
\node[decision] (lpvlpw2) at (0,-16.5) {$l_{\phi V} + l_{\phi W}$};
\node[decision] (lpulpglpx) at (0,-18) {$l_{\phi U}^2 + 4l_{\phi G}l_{\phi X}$};
\node[result,noppn] (nominkbg) at (4.5,0) {no Minkowski background};
\node[result,maybe] (nonconst) at (4.5,-1.5) {non-constant PPN parameters};
\node[result,degen] (nonewton) at (4.5,-3) {no Newtonian limit};
\node[result,good] (rank5a) at (-4,-4.5) {$\beta = \gamma = 1$};
\node[result,good] (rank5b) at (-4,-6) {$\beta = \gamma = 1$};
\node[result,good] (rank4gr) at (7,-7.5) {$\beta = \gamma = 1$};
\node[result,degen] (rank4nsa) at (7,-9) {no $\mathcal{O}(4)$ solution};
\node[result,degen] (rank4nsb) at (7,-10.5) {no $\mathcal{O}(4)$ solution};
\node[result,maybe] (rank4nt) at (3,-12) {$\beta \neq 1, \gamma \neq 1$};
\node[result,degen] (rank3ns) at (-4,-9) {no $\mathcal{O}(2)$ solution};
\node[result,good] (rank3g) at (-4,-10.5) {$\beta = \gamma = 1$};
\node[result,degen] (rank3x0) at (-4,-12) {no $\mathcal{O}(4)$ solution};
\node[result,degen] (rank3xs) at (-4,-13.5) {no $\mathcal{O}(4)$ solution};
\node[result,degen] (rank3na) at (-4,-15) {no $\mathcal{O}(4)$ solution};
\node[result,degen] (rank3nb) at (-4,-16.5) {no $\mathcal{O}(4)$ solution};
\node[result,degen] (rank3nc) at (-4,-18) {no $\mathcal{O}(4)$ solution};
\node[result,good] (rank3y) at (4,-18) {$\beta = \gamma = 1$};
\path (0,1) edge[ultra thick] (l0lp);
\path (l0lp) edge node[above, near start] {$\neq 0$} node[below, near start] {\ref{it:nominkbg}} (nominkbg);
\path (l0lp) edge[ultra thick] node[right, near start] {$= 0$} (lpplppp);
\path (lpplppp) edge node[above, near start] {$\neq 0$} node[below, near start] {\ref{it:nonconst}} (nonconst);
\path (lpplppp) edge[ultra thick] node[right, near start] {$= 0$} (lg);
\path (lg) edge node[above, near start] {$= 0$} node[below, near start] {\ref{it:nonewton}} (nonewton);
\path (lg) edge[ultra thick] node[right, near start] {$\neq 0$} (lulv);
\path (lulv) edge node[above, near start] {$\neq 0$} node[below, near start] {\ref{it:rank5}} (rank5a);
\path (lulv) edge[ultra thick] node[right, near start] {$= 0$} (lvlw);
\path (lvlw) edge node[above, near start] {$\neq 0$} node[below, near start] {\ref{it:rank5}} (rank5b);
\path (lvlw) edge[ultra thick] node[right, near start] {$= 0$} (lulglx);
\path (lulglx) edge node[above, near start] {$\neq 0$} node[below, near start] {\ref{it:rank4}} (lu);
\path (lulglx) edge[ultra thick] node[right, near start] {$= 0$} (lx);
\path (lu) edge node[above, near start] {$= 0$} node[below, near start] {\ref{it:rank4gr}} (rank4gr);
\path (lu) edge node[right, near start] {$\neq 0$} (lpulpv);
\path (lpulpv) edge node[above, near start] {$\neq 0$} node[below, near start] {\ref{it:rank4ns}} (rank4nsa);
\path (lpulpv) edge node[right, near start] {$= 0$} (lpvlpw);
\path (lpvlpw) edge node[above, near start] {$\neq 0$} node[below, near start] {\ref{it:rank4ns}} (rank4nsb);
\path (lpvlpw) edge node[right, near start] {$= 0$} node[left, near start] {\ref{it:rank4nt}} (rank4nt);
\path (lx) edge node[above, near start] {$\neq 0$} node[below, near start] {\ref{it:rank3ns}} (rank3ns);
\path (lx) edge[ultra thick] node[right, near start] {$= 0$} node[left, near start] {\ref{it:rank3}} (lpg);
\path (lpg) edge[ultra thick] node[above, near start] {$= 0$} node[below, near start] {\ref{it:rank3g}} (rank3g);
\path (lpg) edge node[right, near start] {$\neq 0$} (lpx);
\path (lpx) edge node[above, near start] {$= 0$} node[below, near start] {\ref{it:rank3x0}} (rank3x0);
\path (lpx) edge node[right, near start] {$\neq 0$} (lpglpx);
\path (lpglpx) edge node[above, near start] {$> 0$} node[below, near start] {\ref{it:rank3xs}} (rank3xs);
\path (lpglpx) edge node[right, near start] {$< 0$} (lpulpv2);
\path (lpulpv2) edge node[above, near start] {$\neq 0$} node[below, near start] {\ref{it:rank3n}} (rank3na);
\path (lpulpv2) edge node[right, near start] {$= 0$} (lpvlpw2);
\path (lpvlpw2) edge node[above, near start] {$\neq 0$} node[below, near start] {\ref{it:rank3n}} (rank3nb);
\path (lpvlpw2) edge node[right, near start] {$= 0$} (lpulpglpx);
\path (lpulpglpx) edge node[above, near start] {$\neq 0$} node[below, near start] {\ref{it:rank3n}} (rank3nc);
\path (lpulpglpx) edge node[above, near start] {$= 0$} node[below, near start] {\ref{it:rank3y}} (rank3y);
\end{tikzpicture}
\caption{Full classification of scalar-teleparallel theories. The path highlighted by thick arrows corresponds to GTEGR. Theories with \(\beta = \gamma = 1\) are in full agreement with observations. Theories with deviating, but constant PPN parameters receive bounds on their parameters, and are still in agreement if these bounds are met. Theories with massive scalar fields possess distance-dependent PPN parameters and need a more thorough treatment. Other classes of theories are either pathological or need an extension to the standard PPN formalism.}
\label{fig:classes}
\end{figure}
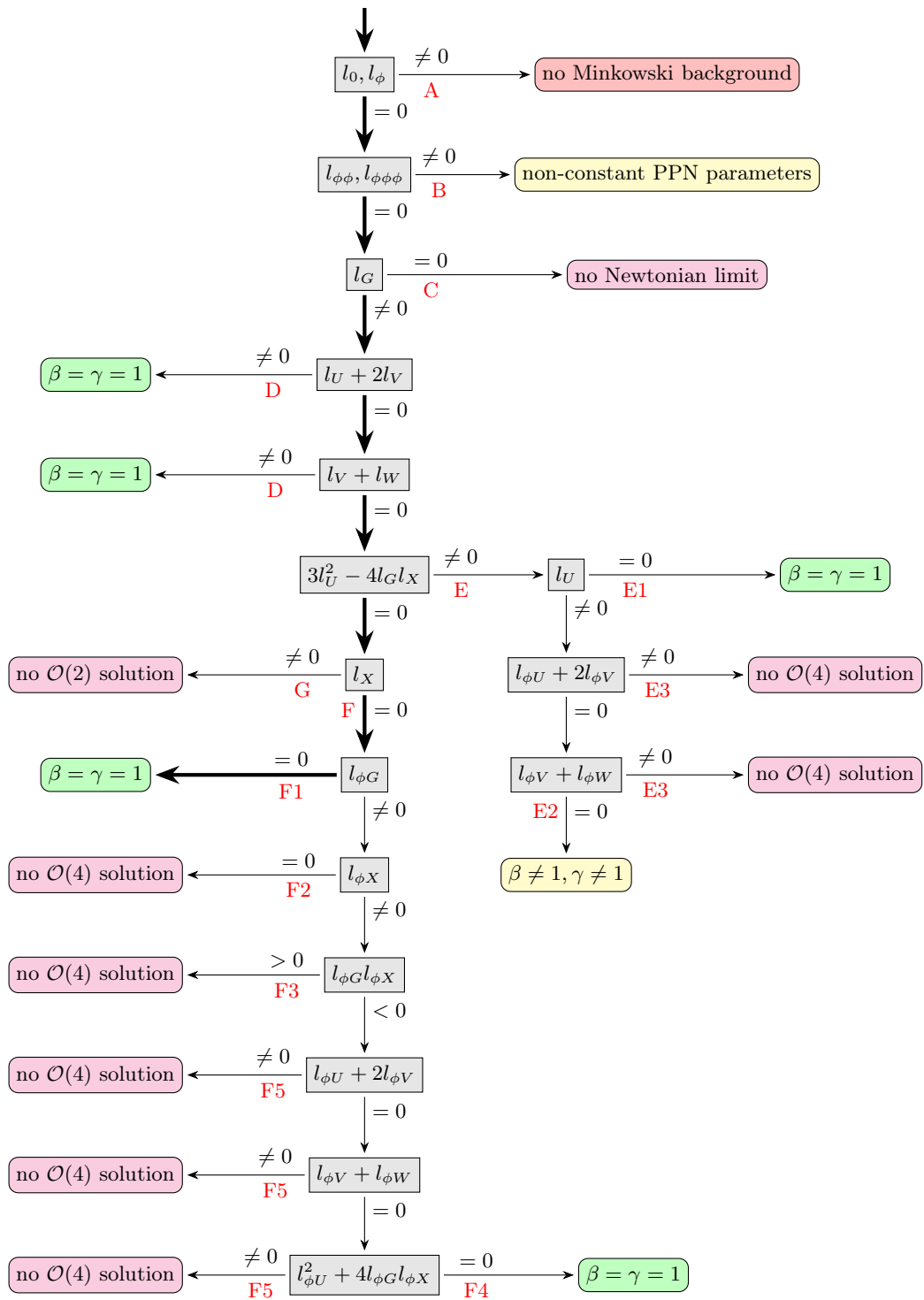

The results gathered in the previous sections, where we solved the perturbed field equations and determined the PPN parameters, can now be used for a complete classification of scalar-teleparallel gravity theories. Here we present this classification, which also serves as a summary of our results. This list is to be understood such that if in one item a particular condition on the Taylor coefficients is assumed to hold, then it will be assumed to be false in all subsequent items.
\begin{enumerate}[A:]
\item\label{it:nominkbg}
If \(l \neq 0\) or \(l_{\phi} \neq 0\), then the zeroth order field equations~\eqref{eq:order0} are not solved by a flat Minkowski background and a constant scalar field, and so the PPN formalism cannot be applied without modifications. In all following cases, we therefore consider \(l = l_{\phi} = 0\).

\item\label{it:nonconst}
If \(l_{\phi\phi} \neq 0\) or \(l_{\phi\phi\phi} \neq 0\), the perturbed field equations contain mass terms for the scalar field, which leads to a Yukawa-type solution, and results in distance-dependent PPN parameters. While such solutions are not excluded, they are not covered by the standard PPN formalism, and thus not considered here.

\item\label{it:nonewton}
If \(l_G = 0\), one finds that the second order field equations yield a contradiction, and so no Newtonian limit exists.

\item\label{it:rank5}
In the case that at least one of \(l_U + 2l_V\) or \(l_V + l_W\) is non-vanishing, which we denoted as rank 5 case, one finds that the solution for the metric perturbations is identical to GR, and thus in particular has \(\beta = \gamma = 1\).

\item\label{it:rank4}
For theories with \(3l_U^2 - 4l_Gl_X \neq 0\) we find that the matrix \(\mat{M}\) in section~\ref{ssec:order2} is of rank 4, and we distinguish the following cases:

\begin{enumerate}[(1)]
\item\label{it:rank4gr}
If \(l_U = 0\), the PPN parameters take the GR values \(\beta = \gamma = 1\).

\item\label{it:rank4nt}
If \(l_{\phi U} + 2l_{\phi V} = l_{\phi V} + l_{\phi W} = 0\), the PPN parameters \(\beta\) and \(\gamma\) take the values~\eqref{eq:ppnparnt}, and thus potentially deviate from their GR values.

\item\label{it:rank4ns}
Otherwise, no solution to the fourth order equations exists.
\end{enumerate}

\item\label{it:rank3}
In the remaining cases \(\mat{M}\) has rank 3. If all of the Taylor coefficients \(l_U, l_V, l_W, l_X\) vanish, we have either no solution to the fourth order equations, or the PPN parameters take the values \(\beta = \gamma = 1\), as given by the following cases:

\begin{enumerate}[(1)]
\item\label{it:rank3g}
If \(l_{\phi G} = 0\), then the fourth order equations can always be solved.

\item\label{it:rank3x0}
If \(l_{\phi X} = 0\), the fourth order equations are contradictory, and no solution exists.

\item\label{it:rank3xs}
If \(l_{\phi X}\) and \(l_{\phi G}\) have the same sign, the scalar field equation can only be solved by an imaginary scalar field, which contradicts the assumption of a real scalar field. Hence, no solution exists.

\item\label{it:rank3y}
If \(l_{\phi U} + 2l_{\phi V} = l_{\phi V} + l_{\phi W} = 0\) and \(l_{\phi U}^2 = -4l_{\phi X}l_{\phi G}\), then the equations can be solved.

\item\label{it:rank3n}
Otherwise, no solution exists.
\end{enumerate}

\item\label{it:rank3ns}
Finally, in all other cases, the second order equations contradict each other, and no solution exists.
\end{enumerate}
The classification is visualized in figure~\ref{fig:classes}. We find numerous cases in which the PPN parameters obtain the GR values \(\beta = \gamma = 1\), and thus pass all experimental tests of post-Newtonian gravity performed thus far. Theories with constant, but deviating PPN parameters may pass these tests as well, provided that their values satisfy the bounds mentioned in the preceding section. Also theories with non-constant PPN parameters may pass these bounds, but require further investigation beyond the standard PPN formalism in order to assess their viability. Theories whose background solution is not a Minkowski background possess a cosmological constant; if its value is of the order of magnitude required to explain the observed late-time accelerating expansion of the universe, it can be neglected on the scales relevant for the PPN formalism, but strongly deviating values are excluded. Theories which do not admit solutions are pathological.

\section{Example theories}\label{sec:examples}
We now apply our findings to a number of example theories. In section~\ref{ssec:stg}, we study a class of theories which is constructed similarly to a well-studied class of scalar-curvature theories of gravity, and contains their general teleparallel equivalent as a particular case. In section~\ref{ssec:noder} we study a simple class of theories with no derivative coupling between the scalar field and the general teleparallel geometry.

\subsection{Scalar-teleparallel analogue of scalar-curvature gravity}\label{ssec:stg}
\begin{figure}[!htbp]
\tikzsetfigurename{bndcoup}
\begin{tikzpicture}[decision/.style={draw,rectangle,fill=black!10!white,outer sep=2pt},result/.style={draw,rectangle,rounded corners,outer sep=2pt},every edge/.style={draw,-Stealth},good/.style={fill=green!25!white},maybe/.style={fill=yellow!25!white},noppn/.style={fill=red!25!white},degen/.style={fill=magenta!25!white}]
\node[decision] (l0lp) at (0,0) {$V, V'$};
\node[decision] (lpplppp) at (0,-1.5) {$V'',V'''$};
\node[decision] (lg) at (0,-3) {$A$};
\node[decision] (lulv) at (0,-4.5) {$C$};
\node[decision] (lglxlu) at (3,-4.5) {$2AB + 3C^2$};
\node[decision] (lx) at (0,-6) {$B$};
\node[decision] (lpg) at (0,-7.5) {$A'$};
\node[decision] (lpx) at (0,-9) {$B'$};
\node[decision] (lpglpx) at (3,-9) {$A'B'$};
\node[decision] (lpulpglpx) at (6,-9) {$2A'B' - C'^2$};
\node[result,noppn] (nominkbg) at (4,0) {no Minkowski background};
\node[result,maybe] (nonconst) at (4,-1.5) {non-constant PPN parameters};
\node[result,degen] (nonewton) at (4,-3) {no Newtonian limit};
\node[result,degen] (o2ns) at (6.5,-4.5) {no $\mathcal{O}(2)$ solution};
\node[result,maybe] (ppnnt) at (3,-6) {$\beta \neq 1, \gamma \neq 1$};
\node[result,good] (ppngr) at (3,-7.5) {$\beta = \gamma = 1$};
\node[result,degen] (o4ns) at (3,-10.5) {no $\mathcal{O}(4)$ solution};
\path (0,1) edge[ultra thick] (l0lp);
\path (l0lp) edge node[above, near start] {$\neq 0$} node[below, near start] {\ref{it:nominkbg}} (nominkbg);
\path (l0lp) edge[ultra thick] node[right, near start] {$= 0$} (lpplppp);
\path (lpplppp) edge node[above, near start] {$\neq 0$} node[below, near start] {\ref{it:nonconst}} (nonconst);
\path (lpplppp) edge[ultra thick] node[right, near start] {$= 0$} (lg);
\path (lg) edge node[above, near start] {$= 0$} node[below, near start] {\ref{it:nonewton}} (nonewton);
\path (lg) edge[ultra thick] node[right, near start] {$\neq 0$} (lulv);
\path (lulv) edge[ultra thick] node[right, near start] {$= 0$} (lx);
\path (lulv) edge node[above, near start] {$\neq 0$} (lglxlu);
\path (lglxlu) edge node[right, near start] {$\neq 0$} node[left, near start] {\ref{it:rank4nt}} (ppnnt);
\path (lglxlu) edge node[above, near start] {$= 0$} node[below, near start] {\ref{it:rank3ns}} (o2ns);
\path (lx) edge node[above right, midway] {$\neq 0$} node[below left, midway] {\ref{it:rank4gr}} (ppngr);
\path (lx) edge[ultra thick] node[right, near start] {$= 0$} (lpg);
\path (lpg) edge[ultra thick] node[above, near start] {$= 0$} node[below, near start] {\ref{it:rank3g}} (ppngr);
\path (lpg) edge node[right, near start] {$\neq 0$} (lpx);
\path (lpx) edge node[above, near start] {$\neq 0$} (lpglpx);
\path (lpx) edge node[above right, midway] {$= 0$} node[below left, midway] {\ref{it:rank3x0}} (o4ns);
\path (lpglpx) edge node[above, near start] {$> 0$} (lpulpglpx);
\path (lpglpx) edge node[right, near start] {$< 0$} node[left, near start] {\ref{it:rank3xs}} (o4ns);
\path (lpulpglpx) edge node[above right, midway] {$= 0$} node[below left, midway] {\ref{it:rank3y}} (ppngr);
\path (lpulpglpx) edge node[above left, midway] {$\neq 0$} node[below right, midway] {\ref{it:rank3n}} (o4ns);
\end{tikzpicture}
\caption{Classification of scalar-teleparallel theories with exclusive coupling to the boundary term.}
\label{fig:bndcoup}
\end{figure}
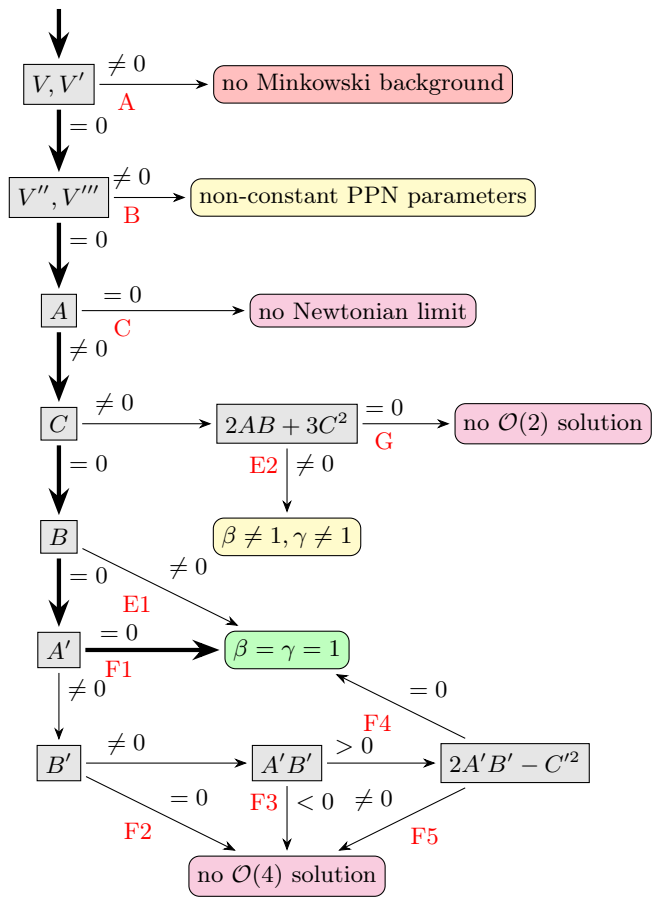

As the first example, we will study a class of theories whose general Lagrangian is of the form
\begin{equation}
L(G, X, U, V, W, \phi) = \mathcal{A}(\phi)G - 2\mathcal{B}(\phi)X - 2\mathcal{C}(\phi)U - 2\mathcal{D}(\phi)V - 2\mathcal{E}(\phi)W + 2\kappa^2\mathcal{V}(\phi)\,,
\end{equation}
with free functions \(\mathcal{A}, \mathcal{B}, \mathcal{C}, \mathcal{D}, \mathcal{E}, \mathcal{V}\) of the scalar field. This Lagrangian generalizes a recently proposed class of scalar-teleparallel gravity theories~\cite{Hohmann:2022mlc}, which is constructed similarly to a well-known class of scalar-curvature theories and contains their teleparallel equivalent as a special case~\cite{Faraoni:2004pi,Fujii:2003pa}. To derive the Taylor coefficients which are required for calculating the PPN parameters, we introduce the notation
\begin{equation}
\mathcal{A}(\Phi) = A\,, \quad
\mathcal{A}'(\Phi) = A'\,, \quad
\mathcal{A}''(\Phi) = A''\,, \quad
\ldots
\end{equation}
and analogously for the remaining parameter functions, for their Taylor coefficients at the cosmological background value \(\Phi\) of the scalar field \(\phi\). With this notation, we can write the Taylor coefficients appearing in the classification in figure~\ref{fig:classes} as
\begin{gather}
l_0 = 2\kappa^2V\,, \quad
l_{\phi} = 2\kappa^2V'\,, \quad
l_{\phi\phi} = 2\kappa^2V''\,, \quad
l_{\phi\phi\phi} = 2\kappa^2V'''\,,\nonumber\\
l_G = A\,, \quad
l_X = -2B\,, \quad
l_U = -2C\,, \quad
l_V = -2D\,, \quad
l_W = -2E\,,\\
l_{\phi G} = A'\,, \quad
l_{\phi X} = -2B'\,, \quad
l_{\phi U} = -2C'\,, \quad
l_{\phi V} = -2D'\,, \quad
l_{\phi W} = -2E'\,.\nonumber
\end{gather}
We see that all Taylor coefficients of the Lagrangian \(L\) depend on independent Taylor coefficients of the parameter functions, and thus all possible cases given in the classification in section~\ref{sec:classes} can be obtained. If we restrict ourselves to the previously proposed class~\cite{Hohmann:2022mlc}
\begin{equation}
L(G, X, U, V, W, \phi) = \mathcal{A}(\phi)G - 2\mathcal{B}(\phi)X - \mathcal{C}(\phi)(2U - V + W) + 2\kappa^2\mathcal{V}(\phi)\,,
\end{equation}
which is motivated by a coupling to the boundary term~\eqref{eq:gtegrbndterms} only, we find that the parameter functions \(\mathcal{D}\) and \(\mathcal{E}\) are bound by the restriction
\begin{equation}
\mathcal{C}(\phi) = -2\mathcal{D}(\phi) = 2\mathcal{E}(\phi)\,.
\end{equation}
Hence, also the corresponding Taylor coefficients satisfy
\begin{equation}
C = -2D = 2E\,, \quad
C' = -2D' = 2E'\,,
\end{equation}
from which then follows
\begin{equation}
l_U = -2l_V = 2l_W\,, \quad
l_{\phi U} = -2l_{\phi V} = 2l_{\phi W}\,.
\end{equation}
By comparison with the classification given in figure~\ref{fig:classes}, we now see that the branches~\ref{it:rank5} and~\ref{it:rank4ns} are excluded for this restricted class of theories, while all other branches can still be obtained by a suitable choice of the remaining Taylor coefficients. We have summarized these remaining cases in figure~\ref{fig:bndcoup}.

\begin{figure}[!htbp]
\tikzsetfigurename{stgequiv}
\begin{tikzpicture}[decision/.style={draw,rectangle,fill=black!10!white,outer sep=2pt},result/.style={draw,rectangle,rounded corners,outer sep=2pt},every edge/.style={draw,-Stealth},good/.style={fill=green!25!white},maybe/.style={fill=yellow!25!white},noppn/.style={fill=red!25!white},degen/.style={fill=magenta!25!white}]
\node[decision] (l0lp) at (0,0) {$V, V'$};
\node[decision] (lpplppp) at (0,-1.5) {$V'',V'''$};
\node[decision] (lg) at (0,-3) {$A$};
\node[decision] (lulv) at (0,-4.5) {$A'$};
\node[decision] (lglxlu) at (3,-4.5) {$2AB + 3A'^2$};
\node[result,noppn] (nominkbg) at (4,0) {no Minkowski background};
\node[result,maybe] (nonconst) at (4,-1.5) {non-constant PPN parameters};
\node[result,degen] (nonewton) at (4,-3) {no Newtonian limit};
\node[result,degen] (o2ns) at (6.5,-4.5) {no $\mathcal{O}(2)$ solution};
\node[result,maybe] (ppnnt) at (3,-6) {$\beta \neq 1, \gamma \neq 1$};
\node[result,good] (ppngr) at (0,-6) {$\beta = \gamma = 1$};
\path (0,1) edge[ultra thick] (l0lp);
\path (l0lp) edge node[above, near start] {$\neq 0$} node[below, near start] {\ref{it:nominkbg}} (nominkbg);
\path (l0lp) edge[ultra thick] node[right, near start] {$= 0$} (lpplppp);
\path (lpplppp) edge node[above, near start] {$\neq 0$} node[below, near start] {\ref{it:nonconst}} (nonconst);
\path (lpplppp) edge[ultra thick] node[right, near start] {$= 0$} (lg);
\path (lg) edge node[above, near start] {$= 0$} node[below, near start] {\ref{it:nonewton}} (nonewton);
\path (lg) edge[ultra thick] node[right, near start] {$\neq 0$} (lulv);
\path (lulv) edge[ultra thick] node[right, near start] {$= 0$} (ppngr);
\path (lulv) edge node[above, near start] {$\neq 0$} (lglxlu);
\path (lglxlu) edge node[right, near start] {$\neq 0$} node[left, near start] {\ref{it:rank4nt}} (ppnnt);
\path (lglxlu) edge node[above, near start] {$= 0$} node[below, near start] {\ref{it:rank3ns}} (o2ns);
\end{tikzpicture}
\caption{Classification of the scalar-teleparallel equivalent of scalar-curvature gravity theories.}
\label{fig:stgequiv}
\end{figure}
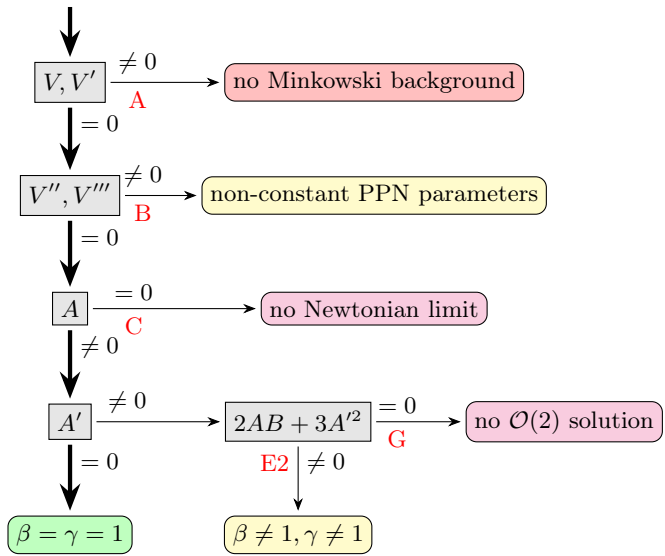

We may further restrict this class to the scalar-teleparallel equivalent of scalar-curvature gravity, which is given by fixing the free function \(\mathcal{C}\) to \(\mathcal{C} = -\mathcal{A}'\), which results in further conditions on the Taylor coefficients given by
\begin{equation}
C = -A'\,, \quad
C' = -A''\,.
\end{equation}
This restriction further excludes the cases~\ref{it:rank3x0}, \ref{it:rank3xs}, \ref{it:rank3y} and~\ref{it:rank3n}, since \(C = 0\) also implies \(A' = 0\), and so the diagram can be simplified to figure~\ref{fig:stgequiv} in this case.

\subsection{Scalar-teleparallel theory without derivative couplings}\label{ssec:noder}
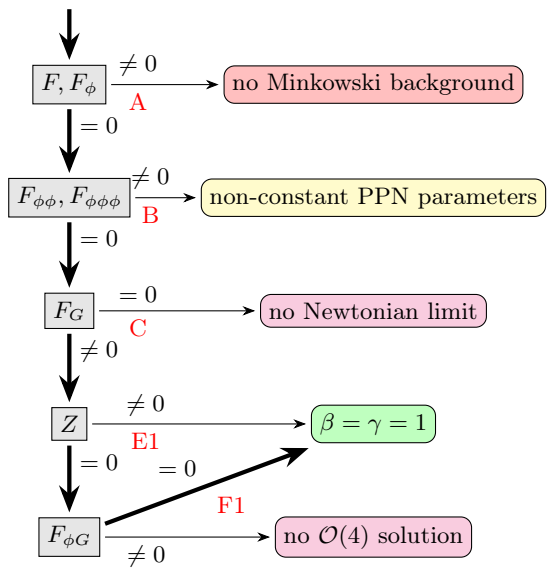
\begin{figure}[!htbp]
\tikzsetfigurename{nodercoup}
\begin{tikzpicture}[decision/.style={draw,rectangle,fill=black!10!white,outer sep=2pt},result/.style={draw,rectangle,rounded corners,outer sep=2pt},every edge/.style={draw,-Stealth},good/.style={fill=green!25!white},maybe/.style={fill=yellow!25!white},noppn/.style={fill=red!25!white},degen/.style={fill=magenta!25!white}]
\node[decision] (l0lp) at (0,0) {$F, F_{\phi}$};
\node[decision] (lpplppp) at (0,-1.5) {$F_{\phi\phi}, F_{\phi\phi\phi}$};
\node[decision] (lg) at (0,-3) {$F_G$};
\node[decision] (lglxlu) at (0,-4.5) {$Z$};
\node[decision] (lpg) at (0,-6) {$F_{\phi G}$};
\node[result,noppn] (nominkbg) at (4,0) {no Minkowski background};
\node[result,maybe] (nonconst) at (4,-1.5) {non-constant PPN parameters};
\node[result,degen] (nonewton) at (4,-3) {no Newtonian limit};
\node[result,good] (ppngr) at (4,-4.5) {$\beta = \gamma = 1$};
\node[result,degen] (o4ns) at (4,-6) {no $\mathcal{O}(4)$ solution};
\path (0,1) edge[ultra thick] (l0lp);
\path (l0lp) edge node[above, near start] {$\neq 0$} node[below, near start] {\ref{it:nominkbg}} (nominkbg);
\path (l0lp) edge[ultra thick] node[right, near start] {$= 0$} (lpplppp);
\path (lpplppp) edge node[above, near start] {$\neq 0$} node[below, near start] {\ref{it:nonconst}} (nonconst);
\path (lpplppp) edge[ultra thick] node[right, near start] {$= 0$} (lg);
\path (lg) edge node[above, near start] {$= 0$} node[below, near start] {\ref{it:nonewton}} (nonewton);
\path (lg) edge[ultra thick] node[right, near start] {$\neq 0$} (lglxlu);
\path (lglxlu) edge[ultra thick] node[right, near start] {$= 0$} (lpg);
\path (lglxlu) edge node[above, near start] {$\neq 0$} node[below, near start] {\ref{it:rank4gr}} (ppngr);
\path (lpg) edge[ultra thick] node[above left, midway] {$= 0$} node[below right, midway] {\ref{it:rank3g}} (ppngr);
\path (lpg) edge node[below, near start] {$\neq 0$} (o4ns);
\end{tikzpicture}
\caption{Classification of scalar-teleparallel theories without derivative couplings.}
\label{fig:nodercoup}
\end{figure}

As another example, we study a class of theories with no derivative coupling to the terms \(U, V, W\), and a simple kinetic term for the scalar field, so that its action takes the form
\begin{equation}\label{eq:ndactiong}
L(G, X, U, V, W, \phi) = \mathcal{F}(G,\phi) - 2\mathcal{Z}(\phi)X\,,
\end{equation}
similarly to a previously studied class of scalar-torsion theories~\cite{Hohmann:2018rwf}. In this case the relevant Taylor coefficients read
\begin{gather}
l_0 = F\,, \quad
l_{\phi} = F_{\phi}\,, \quad
l_{\phi\phi} = F_{\phi\phi}\,, \quad
l_{\phi\phi\phi} = F_{\phi\phi\phi}\,,\nonumber\\
l_G = F_G\,, \quad
l_X = -2Z\,, \quad
l_U = 0\,, \quad
l_V = 0\,, \quad
l_W = 0\,,\\
l_{\phi G} = F_{\phi G}\,, \quad
l_{\phi X} = -2Z_{\phi}\,, \quad
l_{\phi U} = 0\,, \quad
l_{\phi V} = 0\,, \quad
l_{\phi W} = 0\,.\nonumber
\end{gather}
The vanishing derivative coupling terms exclude most of the branches we have found in section~\ref{sec:classes}. The only remaining branches are~\ref{it:nominkbg}, \ref{it:nonconst}, \ref{it:nonewton}, \ref{it:rank4gr}, \ref{it:rank3g}, \ref{it:rank3x0}, \ref{it:rank3xs} and \ref{it:rank3n}. They are summarized in figure~\ref{fig:nodercoup}.

\section{Conclusion}\label{sec:conclusion}
In this article we have studied the post-Newtonian limit of a general class of scalar-teleparallel theories of gravity and obtained a full classification of these theories based on the existence of a post-Newtonian solution, its uniqueness, as well as the values of their PPN parameters. Our findings allow to exclude various classes of theories, since they do not allow for a post-Newtonian solution to their field equations. For the remaining cases, for which a post-Newtonian solution exists and all PPN parameters can be determined, we find that these theories are always fully conservative, which means that there are no preferred frame or preferred location effects, and no violation of the conservation of total momentum. The only PPN parameters which may possibly deviate from their GR values in such theories are \(\beta\) and \(\gamma\). Within this class of scalar-teleparallel theories, we found one subclass in which both \(\beta\) and \(\gamma\) deviate from the GR values, so that it receives bounds from observations. We also found numerous classes for which the PPN parameters take the GR values \(\beta = \gamma = 1\), and so they are in full agreement with experimental tests of these and the remaining PPN parameters, and indistinguishable from GR by such observations.

Among the theories we have studied we also found several cases which require further investigation. This includes in particular the case in which the scalar field obtains a non-vanishing mass, so that its post-Newtonian solution is not given by a Newtonian potential, but by a Yukawa-type potential, which then results in distance-dependent PPN parameters. Theories of this type are not covered by the standard PPN formalism and require a more detailed calculation in order to compare them to observations~\cite{Deng:2016moh,Hohmann:2017qje}. Another class of theories which requires further studies are those for which no unique solution to the perturbed field equations is obtained, which is apparent by the fact that the connection perturbation is undetermined by the perturbed field equations at the same perturbation order. Theories of this type may suffer from a strong coupling problem, which has been found in other classes of teleparallel gravity theories, and is an actively debated topic~\cite{Golovnev:2018wbh,Golovnev:2020zpv,BeltranJimenez:2020fvy,Blagojevic:2020dyq,Guzman:2019oth,BeltranJimenez:2019nns,Bahamonde:2022ohm,Golovnev:2020nln,Li:2023fto}.

Our work further invites for the study of more general classes of theories within the field of general teleparallel gravity. The most straightforward generalization is the construction of a theory including more general derivative couplings, as motivated by the Horndeski class of gravity theories~\cite{Horndeski:1974wa,Kobayashi:2011nu,Kobayashi:2019hrl}, which has been done for the cases of scalar-torsion gravity~\cite{Bahamonde:2019shr,Bahamonde:2019ipm,Bahamonde:2020cfv} and scalar-nonmetricity gravity~\cite{Bahamonde:2022cmz}. Another class of theories worth studying is the general teleparallel quadratic gravity~\cite{BeltranJimenez:2019odq}, whose PPN parameters could be calculated similarly to the known cases of new general relativity~\cite{Ualikhanova:2019ygl} and newer general relativity~\cite{Flathmann:2021khc}.

\begin{acknowledgments}
This research was funded by the Science Committee of the Ministry of Science and Higher Education of the Republic of Kazakhstan (Grant No. AP14972654).
MH gratefully acknowledges the full financial support by the Estonian Research Council through the Personal Research Funding project PRG356.
\end{acknowledgments}

\bibliography{teleppn}

\begin{thebibliography}{68}%
\makeatletter
\providecommand \@ifxundefined [1]{%
 \@ifx{#1\undefined}
}%
\providecommand \@ifnum [1]{%
 \ifnum #1\expandafter \@firstoftwo
 \else \expandafter \@secondoftwo
 \fi
}%
\providecommand \@ifx [1]{%
 \ifx #1\expandafter \@firstoftwo
 \else \expandafter \@secondoftwo
 \fi
}%
\providecommand \natexlab [1]{#1}%
\providecommand \enquote  [1]{``#1''}%
\providecommand \bibnamefont  [1]{#1}%
\providecommand \bibfnamefont [1]{#1}%
\providecommand \citenamefont [1]{#1}%
\providecommand \href@noop [0]{\@secondoftwo}%
\providecommand \href [0]{\begingroup \@sanitize@url \@href}%
\providecommand \@href[1]{\@@startlink{#1}\@@href}%
\providecommand \@@href[1]{\endgroup#1\@@endlink}%
\providecommand \@sanitize@url [0]{\catcode `\\12\catcode `\$12\catcode
  `\&12\catcode `\#12\catcode `\^12\catcode `\_12\catcode `\%12\relax}%
\providecommand \@@startlink[1]{}%
\providecommand \@@endlink[0]{}%
\providecommand \url  [0]{\begingroup\@sanitize@url \@url }%
\providecommand \@url [1]{\endgroup\@href {#1}{\urlprefix }}%
\providecommand \urlprefix  [0]{URL }%
\providecommand \Eprint [0]{\href }%
\providecommand \doibase [0]{https://doi.org/}%
\providecommand \selectlanguage [0]{\@gobble}%
\providecommand \bibinfo  [0]{\@secondoftwo}%
\providecommand \bibfield  [0]{\@secondoftwo}%
\providecommand \translation [1]{[#1]}%
\providecommand \BibitemOpen [0]{}%
\providecommand \bibitemStop [0]{}%
\providecommand \bibitemNoStop [0]{.\EOS\space}%
\providecommand \EOS [0]{\spacefactor3000\relax}%
\providecommand \BibitemShut  [1]{\csname bibitem#1\endcsname}%
\let\auto@bib@innerbib\@empty
\bibitem [{\citenamefont {Gillessen}\ \emph {et~al.}(2009)\citenamefont
  {Gillessen}, \citenamefont {Eisenhauer}, \citenamefont {Trippe},
  \citenamefont {Alexander}, \citenamefont {Genzel}, \citenamefont {Martins},\
  and\ \citenamefont {Ott}}]{Gillessen:2008qv}%
  \BibitemOpen
  \bibfield  {author} {\bibinfo {author} {\bibfnamefont {S.}~\bibnamefont
  {Gillessen}}, \bibinfo {author} {\bibfnamefont {F.}~\bibnamefont
  {Eisenhauer}}, \bibinfo {author} {\bibfnamefont {S.}~\bibnamefont {Trippe}},
  \bibinfo {author} {\bibfnamefont {T.}~\bibnamefont {Alexander}}, \bibinfo
  {author} {\bibfnamefont {R.}~\bibnamefont {Genzel}}, \bibinfo {author}
  {\bibfnamefont {F.}~\bibnamefont {Martins}},\ and\ \bibinfo {author}
  {\bibfnamefont {T.}~\bibnamefont {Ott}},\ }\bibfield  {title} {\bibinfo
  {title} {{Monitoring stellar orbits around the Massive Black Hole in the
  Galactic Center}},\ }\href {https://doi.org/10.1088/0004-637X/692/2/1075}
  {\bibfield  {journal} {\bibinfo  {journal} {Astrophys. J.}\ }\textbf
  {\bibinfo {volume} {692}},\ \bibinfo {pages} {1075} (\bibinfo {year}
  {2009})},\ \Eprint {https://arxiv.org/abs/0810.4674} {arXiv:0810.4674
  [astro-ph]} \BibitemShut {NoStop}%
\bibitem [{\citenamefont {Abbott}\ \emph {et~al.}(2016)\citenamefont {Abbott}
  \emph {et~al.}}]{LIGOScientific:2016aoc}%
  \BibitemOpen
  \bibfield  {author} {\bibinfo {author} {\bibfnamefont {B.~P.}\ \bibnamefont
  {Abbott}} \emph {et~al.} (\bibinfo {collaboration} {LIGO Scientific,
  Virgo}),\ }\bibfield  {title} {\bibinfo {title} {{Observation of
  Gravitational Waves from a Binary Black Hole Merger}},\ }\href
  {https://doi.org/10.1103/PhysRevLett.116.061102} {\bibfield  {journal}
  {\bibinfo  {journal} {Phys. Rev. Lett.}\ }\textbf {\bibinfo {volume} {116}},\
  \bibinfo {pages} {061102} (\bibinfo {year} {2016})},\ \Eprint
  {https://arxiv.org/abs/1602.03837} {arXiv:1602.03837 [gr-qc]} \BibitemShut
  {NoStop}%
\bibitem [{\citenamefont {Akiyama}\ \emph {et~al.}(2019)\citenamefont {Akiyama}
  \emph {et~al.}}]{EventHorizonTelescope:2019dse}%
  \BibitemOpen
  \bibfield  {author} {\bibinfo {author} {\bibfnamefont {K.}~\bibnamefont
  {Akiyama}} \emph {et~al.} (\bibinfo {collaboration} {Event Horizon
  Telescope}),\ }\bibfield  {title} {\bibinfo {title} {{First M87 Event Horizon
  Telescope Results. I. The Shadow of the Supermassive Black Hole}},\ }\href
  {https://doi.org/10.3847/2041-8213/ab0ec7} {\bibfield  {journal} {\bibinfo
  {journal} {Astrophys. J. Lett.}\ }\textbf {\bibinfo {volume} {875}},\
  \bibinfo {pages} {L1} (\bibinfo {year} {2019})},\ \Eprint
  {https://arxiv.org/abs/1906.11238} {arXiv:1906.11238 [astro-ph.GA]}
  \BibitemShut {NoStop}%
\bibitem [{\citenamefont {Akiyama}\ \emph {et~al.}(2022)\citenamefont {Akiyama}
  \emph {et~al.}}]{EventHorizonTelescope:2022wkp}%
  \BibitemOpen
  \bibfield  {author} {\bibinfo {author} {\bibfnamefont {K.}~\bibnamefont
  {Akiyama}} \emph {et~al.} (\bibinfo {collaboration} {Event Horizon
  Telescope}),\ }\bibfield  {title} {\bibinfo {title} {{First Sagittarius A*
  Event Horizon Telescope Results. I. The Shadow of the Supermassive Black Hole
  in the Center of the Milky Way}},\ }\href
  {https://doi.org/10.3847/2041-8213/ac6674} {\bibfield  {journal} {\bibinfo
  {journal} {Astrophys. J. Lett.}\ }\textbf {\bibinfo {volume} {930}},\
  \bibinfo {pages} {L12} (\bibinfo {year} {2022})}\BibitemShut {NoStop}%
\bibitem [{\citenamefont {Aghanim}\ \emph {et~al.}(2020)\citenamefont {Aghanim}
  \emph {et~al.}}]{Planck:2018vyg}%
  \BibitemOpen
  \bibfield  {author} {\bibinfo {author} {\bibfnamefont {N.}~\bibnamefont
  {Aghanim}} \emph {et~al.} (\bibinfo {collaboration} {Planck}),\ }\bibfield
  {title} {\bibinfo {title} {{Planck 2018 results. VI. Cosmological
  parameters}},\ }\href {https://doi.org/10.1051/0004-6361/201833910}
  {\bibfield  {journal} {\bibinfo  {journal} {Astron. Astrophys.}\ }\textbf
  {\bibinfo {volume} {641}},\ \bibinfo {pages} {A6} (\bibinfo {year} {2020})},\
  \bibinfo {note} {[Erratum: Astron.Astrophys. 652, C4 (2021)]},\ \Eprint
  {https://arxiv.org/abs/1807.06209} {arXiv:1807.06209 [astro-ph.CO]}
  \BibitemShut {NoStop}%
\bibitem [{\citenamefont {Di~Valentino}\ \emph {et~al.}(2021)\citenamefont
  {Di~Valentino}, \citenamefont {Mena}, \citenamefont {Pan}, \citenamefont
  {Visinelli}, \citenamefont {Yang}, \citenamefont {Melchiorri}, \citenamefont
  {Mota}, \citenamefont {Riess},\ and\ \citenamefont
  {Silk}}]{DiValentino:2021izs}%
  \BibitemOpen
  \bibfield  {author} {\bibinfo {author} {\bibfnamefont {E.}~\bibnamefont
  {Di~Valentino}}, \bibinfo {author} {\bibfnamefont {O.}~\bibnamefont {Mena}},
  \bibinfo {author} {\bibfnamefont {S.}~\bibnamefont {Pan}}, \bibinfo {author}
  {\bibfnamefont {L.}~\bibnamefont {Visinelli}}, \bibinfo {author}
  {\bibfnamefont {W.}~\bibnamefont {Yang}}, \bibinfo {author} {\bibfnamefont
  {A.}~\bibnamefont {Melchiorri}}, \bibinfo {author} {\bibfnamefont {D.~F.}\
  \bibnamefont {Mota}}, \bibinfo {author} {\bibfnamefont {A.~G.}\ \bibnamefont
  {Riess}},\ and\ \bibinfo {author} {\bibfnamefont {J.}~\bibnamefont {Silk}},\
  }\bibfield  {title} {\bibinfo {title} {{In the realm of the Hubble
  tension\textemdash{}a review of solutions}},\ }\href
  {https://doi.org/10.1088/1361-6382/ac086d} {\bibfield  {journal} {\bibinfo
  {journal} {Class. Quant. Grav.}\ }\textbf {\bibinfo {volume} {38}},\ \bibinfo
  {pages} {153001} (\bibinfo {year} {2021})},\ \Eprint
  {https://arxiv.org/abs/2103.01183} {arXiv:2103.01183 [astro-ph.CO]}
  \BibitemShut {NoStop}%
\bibitem [{\citenamefont {Nojiri}\ and\ \citenamefont
  {Odintsov}(2006)}]{Nojiri:2006ri}%
  \BibitemOpen
  \bibfield  {author} {\bibinfo {author} {\bibfnamefont {S.}~\bibnamefont
  {Nojiri}}\ and\ \bibinfo {author} {\bibfnamefont {S.~D.}\ \bibnamefont
  {Odintsov}},\ }\bibfield  {title} {\bibinfo {title} {{Introduction to
  modified gravity and gravitational alternative for dark energy}},\ }\bibfield
   {booktitle} {\emph {\bibinfo {booktitle} {{Theoretical physics: Current
  mathematical topics in gravitation and cosmology. Proceedings, 42nd Karpacz
  Winter School, Ladek, Poland, February 6-11, 2006}}},\ }\href
  {https://doi.org/10.1142/S0219887807001928} {\bibfield  {journal} {\bibinfo
  {journal} {eConf}\ }\textbf {\bibinfo {volume} {C0602061}},\ \bibinfo {pages}
  {06} (\bibinfo {year} {2006})},\ \bibinfo {note} {[Int. J. Geom. Meth. Mod.
  Phys.4,115(2007)]},\ \Eprint {https://arxiv.org/abs/hep-th/0601213}
  {arXiv:hep-th/0601213 [hep-th]} \BibitemShut {NoStop}%
\bibitem [{\citenamefont {Nojiri}\ and\ \citenamefont
  {Odintsov}(2011)}]{Nojiri:2010wj}%
  \BibitemOpen
  \bibfield  {author} {\bibinfo {author} {\bibfnamefont {S.}~\bibnamefont
  {Nojiri}}\ and\ \bibinfo {author} {\bibfnamefont {S.~D.}\ \bibnamefont
  {Odintsov}},\ }\bibfield  {title} {\bibinfo {title} {{Unified cosmic history
  in modified gravity: from F(R) theory to Lorentz non-invariant models}},\
  }\href {https://doi.org/10.1016/j.physrep.2011.04.001} {\bibfield  {journal}
  {\bibinfo  {journal} {Phys. Rept.}\ }\textbf {\bibinfo {volume} {505}},\
  \bibinfo {pages} {59} (\bibinfo {year} {2011})},\ \Eprint
  {https://arxiv.org/abs/1011.0544} {arXiv:1011.0544 [gr-qc]} \BibitemShut
  {NoStop}%
\bibitem [{\citenamefont {Faraoni}\ and\ \citenamefont
  {Capozziello}(2011)}]{Capozziello:2010zz}%
  \BibitemOpen
  \bibfield  {author} {\bibinfo {author} {\bibfnamefont {V.}~\bibnamefont
  {Faraoni}}\ and\ \bibinfo {author} {\bibfnamefont {S.}~\bibnamefont
  {Capozziello}},\ }\href {https://doi.org/10.1007/978-94-007-0165-6} {\emph
  {\bibinfo {title} {{Beyond Einstein Gravity}}}},\ Vol.\ \bibinfo {volume}
  {170}\ (\bibinfo  {publisher} {Springer},\ \bibinfo {address} {Dordrecht},\
  \bibinfo {year} {2011})\BibitemShut {NoStop}%
\bibitem [{\citenamefont {Clifton}\ \emph {et~al.}(2012)\citenamefont
  {Clifton}, \citenamefont {Ferreira}, \citenamefont {Padilla},\ and\
  \citenamefont {Skordis}}]{Clifton:2011jh}%
  \BibitemOpen
  \bibfield  {author} {\bibinfo {author} {\bibfnamefont {T.}~\bibnamefont
  {Clifton}}, \bibinfo {author} {\bibfnamefont {P.~G.}\ \bibnamefont
  {Ferreira}}, \bibinfo {author} {\bibfnamefont {A.}~\bibnamefont {Padilla}},\
  and\ \bibinfo {author} {\bibfnamefont {C.}~\bibnamefont {Skordis}},\
  }\bibfield  {title} {\bibinfo {title} {{Modified Gravity and Cosmology}},\
  }\href {https://doi.org/10.1016/j.physrep.2012.01.001} {\bibfield  {journal}
  {\bibinfo  {journal} {Phys. Rept.}\ }\textbf {\bibinfo {volume} {513}},\
  \bibinfo {pages} {1} (\bibinfo {year} {2012})},\ \Eprint
  {https://arxiv.org/abs/1106.2476} {arXiv:1106.2476 [astro-ph.CO]}
  \BibitemShut {NoStop}%
\bibitem [{\citenamefont {Nojiri}\ \emph {et~al.}(2017)\citenamefont {Nojiri},
  \citenamefont {Odintsov},\ and\ \citenamefont {Oikonomou}}]{Nojiri:2017ncd}%
  \BibitemOpen
  \bibfield  {author} {\bibinfo {author} {\bibfnamefont {S.}~\bibnamefont
  {Nojiri}}, \bibinfo {author} {\bibfnamefont {S.~D.}\ \bibnamefont
  {Odintsov}},\ and\ \bibinfo {author} {\bibfnamefont {V.~K.}\ \bibnamefont
  {Oikonomou}},\ }\bibfield  {title} {\bibinfo {title} {{Modified Gravity
  Theories on a Nutshell: Inflation, Bounce and Late-time Evolution}},\ }\href
  {https://doi.org/10.1016/j.physrep.2017.06.001} {\bibfield  {journal}
  {\bibinfo  {journal} {Phys. Rept.}\ }\textbf {\bibinfo {volume} {692}},\
  \bibinfo {pages} {1} (\bibinfo {year} {2017})},\ \Eprint
  {https://arxiv.org/abs/1705.11098} {arXiv:1705.11098 [gr-qc]} \BibitemShut
  {NoStop}%
\bibitem [{\citenamefont {Bull}\ \emph {et~al.}(2016)\citenamefont {Bull} \emph
  {et~al.}}]{Bull:2015stt}%
  \BibitemOpen
  \bibfield  {author} {\bibinfo {author} {\bibfnamefont {P.}~\bibnamefont
  {Bull}} \emph {et~al.},\ }\bibfield  {title} {\bibinfo {title} {{Beyond
  $\Lambda$CDM: Problems, solutions, and the road ahead}},\ }\href
  {https://doi.org/10.1016/j.dark.2016.02.001} {\bibfield  {journal} {\bibinfo
  {journal} {Phys. Dark Univ.}\ }\textbf {\bibinfo {volume} {12}},\ \bibinfo
  {pages} {56} (\bibinfo {year} {2016})},\ \Eprint
  {https://arxiv.org/abs/1512.05356} {arXiv:1512.05356 [astro-ph.CO]}
  \BibitemShut {NoStop}%
\bibitem [{\citenamefont {Heisenberg}(2019)}]{Heisenberg:2018vsk}%
  \BibitemOpen
  \bibfield  {author} {\bibinfo {author} {\bibfnamefont {L.}~\bibnamefont
  {Heisenberg}},\ }\bibfield  {title} {\bibinfo {title} {{A systematic approach
  to generalisations of General Relativity and their cosmological
  implications}},\ }\href {https://doi.org/10.1016/j.physrep.2018.11.006}
  {\bibfield  {journal} {\bibinfo  {journal} {Phys. Rept.}\ }\textbf {\bibinfo
  {volume} {796}},\ \bibinfo {pages} {1} (\bibinfo {year} {2019})},\ \Eprint
  {https://arxiv.org/abs/1807.01725} {arXiv:1807.01725 [gr-qc]} \BibitemShut
  {NoStop}%
\bibitem [{\citenamefont {Akrami}\ \emph {et~al.}(2021)\citenamefont {Akrami}
  \emph {et~al.}}]{CANTATA:2021ktz}%
  \BibitemOpen
  \bibfield  {author} {\bibinfo {author} {\bibfnamefont {Y.}~\bibnamefont
  {Akrami}} \emph {et~al.} (\bibinfo {collaboration} {CANTATA}),\ }\href
  {https://doi.org/10.1007/978-3-030-83715-0} {\emph {\bibinfo {title}
  {{Modified Gravity and Cosmology. An Update by the CANTATA Network}}}},\
  edited by\ \bibinfo {editor} {\bibfnamefont {E.~N.}\ \bibnamefont
  {Saridakis}}, \bibinfo {editor} {\bibfnamefont {R.}~\bibnamefont {Lazkoz}},
  \bibinfo {editor} {\bibfnamefont {V.}~\bibnamefont {Salzano}}, \bibinfo
  {editor} {\bibfnamefont {P.}~\bibnamefont {Vargas~Moniz}}, \bibinfo {editor}
  {\bibfnamefont {S.}~\bibnamefont {Capozziello}}, \bibinfo {editor}
  {\bibfnamefont {J.}~\bibnamefont {Beltr\'an~Jim\'enez}}, \bibinfo {editor}
  {\bibfnamefont {M.}~\bibnamefont {De~Laurentis}},\ and\ \bibinfo {editor}
  {\bibfnamefont {G.~J.}\ \bibnamefont {Olmo}}\ (\bibinfo  {publisher}
  {Springer},\ \bibinfo {year} {2021})\ \Eprint
  {https://arxiv.org/abs/2105.12582} {arXiv:2105.12582 [gr-qc]} \BibitemShut
  {NoStop}%
\bibitem [{\citenamefont {Capozziello}\ \emph {et~al.}(2022)\citenamefont
  {Capozziello}, \citenamefont {De~Falco},\ and\ \citenamefont
  {Ferrara}}]{Capozziello:2022zzh}%
  \BibitemOpen
  \bibfield  {author} {\bibinfo {author} {\bibfnamefont {S.}~\bibnamefont
  {Capozziello}}, \bibinfo {author} {\bibfnamefont {V.}~\bibnamefont
  {De~Falco}},\ and\ \bibinfo {author} {\bibfnamefont {C.}~\bibnamefont
  {Ferrara}},\ }\bibfield  {title} {\bibinfo {title} {{Comparing equivalent
  gravities: common features and differences}},\ }\href
  {https://doi.org/10.1140/epjc/s10052-022-10823-x} {\bibfield  {journal}
  {\bibinfo  {journal} {Eur. Phys. J. C}\ }\textbf {\bibinfo {volume} {82}},\
  \bibinfo {pages} {865} (\bibinfo {year} {2022})},\ \Eprint
  {https://arxiv.org/abs/2208.03011} {arXiv:2208.03011 [gr-qc]} \BibitemShut
  {NoStop}%
\bibitem [{\citenamefont {Odintsov}\ \emph {et~al.}(2023)\citenamefont
  {Odintsov}, \citenamefont {Oikonomou}, \citenamefont {Giannakoudi},
  \citenamefont {Fronimos},\ and\ \citenamefont
  {Lymperiadou}}]{Odintsov:2023weg}%
  \BibitemOpen
  \bibfield  {author} {\bibinfo {author} {\bibfnamefont {S.~D.}\ \bibnamefont
  {Odintsov}}, \bibinfo {author} {\bibfnamefont {V.~K.}\ \bibnamefont
  {Oikonomou}}, \bibinfo {author} {\bibfnamefont {I.}~\bibnamefont
  {Giannakoudi}}, \bibinfo {author} {\bibfnamefont {F.~P.}\ \bibnamefont
  {Fronimos}},\ and\ \bibinfo {author} {\bibfnamefont {E.~C.}\ \bibnamefont
  {Lymperiadou}},\ }\bibfield  {title} {\bibinfo {title} {{Recent Advances in
  Inflation}},\ }\href {https://doi.org/10.3390/sym15091701} {\bibfield
  {journal} {\bibinfo  {journal} {Symmetry}\ }\textbf {\bibinfo {volume}
  {15}},\ \bibinfo {pages} {1701} (\bibinfo {year} {2023})},\ \Eprint
  {https://arxiv.org/abs/2307.16308} {arXiv:2307.16308 [gr-qc]} \BibitemShut
  {NoStop}%
\bibitem [{\citenamefont {Jiménez}\ \emph {et~al.}(2019)\citenamefont
  {Jiménez}, \citenamefont {Heisenberg},\ and\ \citenamefont
  {Koivisto}}]{BeltranJimenez:2019tjy}%
  \BibitemOpen
  \bibfield  {author} {\bibinfo {author} {\bibfnamefont {J.~B.}\ \bibnamefont
  {Jiménez}}, \bibinfo {author} {\bibfnamefont {L.}~\bibnamefont
  {Heisenberg}},\ and\ \bibinfo {author} {\bibfnamefont {T.~S.}\ \bibnamefont
  {Koivisto}},\ }\bibfield  {title} {\bibinfo {title} {{The Geometrical Trinity
  of Gravity}},\ }\href {https://doi.org/10.3390/universe5070173} {\bibfield
  {journal} {\bibinfo  {journal} {Universe}\ }\textbf {\bibinfo {volume} {5}},\
  \bibinfo {pages} {173} (\bibinfo {year} {2019})},\ \Eprint
  {https://arxiv.org/abs/1903.06830} {arXiv:1903.06830 [hep-th]} \BibitemShut
  {NoStop}%
\bibitem [{\citenamefont {Beltr\'an~Jim\'enez}\ \emph
  {et~al.}(2020)\citenamefont {Beltr\'an~Jim\'enez}, \citenamefont
  {Heisenberg}, \citenamefont {Iosifidis}, \citenamefont {Jim\'enez-Cano},\
  and\ \citenamefont {Koivisto}}]{BeltranJimenez:2019odq}%
  \BibitemOpen
  \bibfield  {author} {\bibinfo {author} {\bibfnamefont {J.}~\bibnamefont
  {Beltr\'an~Jim\'enez}}, \bibinfo {author} {\bibfnamefont {L.}~\bibnamefont
  {Heisenberg}}, \bibinfo {author} {\bibfnamefont {D.}~\bibnamefont
  {Iosifidis}}, \bibinfo {author} {\bibfnamefont {A.}~\bibnamefont
  {Jim\'enez-Cano}},\ and\ \bibinfo {author} {\bibfnamefont {T.~S.}\
  \bibnamefont {Koivisto}},\ }\bibfield  {title} {\bibinfo {title} {{General
  teleparallel quadratic gravity}},\ }\href
  {https://doi.org/10.1016/j.physletb.2020.135422} {\bibfield  {journal}
  {\bibinfo  {journal} {Phys. Lett. B}\ }\textbf {\bibinfo {volume} {805}},\
  \bibinfo {pages} {135422} (\bibinfo {year} {2020})},\ \Eprint
  {https://arxiv.org/abs/1909.09045} {arXiv:1909.09045 [gr-qc]} \BibitemShut
  {NoStop}%
\bibitem [{\citenamefont {Hohmann}(2023)}]{Hohmann:2022mlc}%
  \BibitemOpen
  \bibfield  {author} {\bibinfo {author} {\bibfnamefont {M.}~\bibnamefont
  {Hohmann}},\ }\bibfield  {title} {\bibinfo {title} {{Teleparallel Gravity}},\
  }\href {https://doi.org/10.1007/978-3-031-31520-6_4} {\bibfield  {journal}
  {\bibinfo  {journal} {Lect. Notes Phys.}\ }\textbf {\bibinfo {volume}
  {1017}},\ \bibinfo {pages} {145} (\bibinfo {year} {2023})},\ \Eprint
  {https://arxiv.org/abs/2207.06438} {arXiv:2207.06438 [gr-qc]} \BibitemShut
  {NoStop}%
\bibitem [{\citenamefont {Heisenberg}\ \emph {et~al.}(2023)\citenamefont
  {Heisenberg}, \citenamefont {Hohmann},\ and\ \citenamefont
  {Kuhn}}]{Heisenberg:2022mbo}%
  \BibitemOpen
  \bibfield  {author} {\bibinfo {author} {\bibfnamefont {L.}~\bibnamefont
  {Heisenberg}}, \bibinfo {author} {\bibfnamefont {M.}~\bibnamefont
  {Hohmann}},\ and\ \bibinfo {author} {\bibfnamefont {S.}~\bibnamefont
  {Kuhn}},\ }\bibfield  {title} {\bibinfo {title} {{Homogeneous and isotropic
  cosmology in general teleparallel gravity}},\ }\href
  {https://doi.org/10.1140/epjc/s10052-023-11462-6} {\bibfield  {journal}
  {\bibinfo  {journal} {Eur. Phys. J. C}\ }\textbf {\bibinfo {volume} {83}},\
  \bibinfo {pages} {315} (\bibinfo {year} {2023})},\ \Eprint
  {https://arxiv.org/abs/2212.14324} {arXiv:2212.14324 [gr-qc]} \BibitemShut
  {NoStop}%
\bibitem [{\citenamefont {Heisenberg}\ and\ \citenamefont
  {Hohmann}(2024)}]{Heisenberg:2023tho}%
  \BibitemOpen
  \bibfield  {author} {\bibinfo {author} {\bibfnamefont {L.}~\bibnamefont
  {Heisenberg}}\ and\ \bibinfo {author} {\bibfnamefont {M.}~\bibnamefont
  {Hohmann}},\ }\bibfield  {title} {\bibinfo {title} {{Gauge-invariant
  cosmological perturbations in general teleparallel gravity}},\ }\href
  {https://doi.org/10.1140/epjc/s10052-024-12810-w} {\bibfield  {journal}
  {\bibinfo  {journal} {Eur. Phys. J. C}\ }\textbf {\bibinfo {volume} {84}},\
  \bibinfo {pages} {462} (\bibinfo {year} {2024})},\ \Eprint
  {https://arxiv.org/abs/2311.05597} {arXiv:2311.05597 [gr-qc]} \BibitemShut
  {NoStop}%
\bibitem [{\citenamefont {Heisenberg}\ \emph {et~al.}(2024)\citenamefont
  {Heisenberg}, \citenamefont {Hohmann},\ and\ \citenamefont
  {Kuhn}}]{Heisenberg:2023wgk}%
  \BibitemOpen
  \bibfield  {author} {\bibinfo {author} {\bibfnamefont {L.}~\bibnamefont
  {Heisenberg}}, \bibinfo {author} {\bibfnamefont {M.}~\bibnamefont
  {Hohmann}},\ and\ \bibinfo {author} {\bibfnamefont {S.}~\bibnamefont
  {Kuhn}},\ }\bibfield  {title} {\bibinfo {title} {{Cosmological teleparallel
  perturbations}},\ }\href {https://doi.org/10.1088/1475-7516/2024/03/063}
  {\bibfield  {journal} {\bibinfo  {journal} {JCAP}\ }\textbf {\bibinfo
  {volume} {03}},\ \bibinfo {pages} {063}},\ \Eprint
  {https://arxiv.org/abs/2311.05495} {arXiv:2311.05495 [gr-qc]} \BibitemShut
  {NoStop}%
\bibitem [{\citenamefont {Will}(1993)}]{Will:1993ns}%
  \BibitemOpen
  \bibfield  {author} {\bibinfo {author} {\bibfnamefont {C.~M.}\ \bibnamefont
  {Will}},\ }\href {https://doi.org/10.1017/CBO9780511564246} {\emph {\bibinfo
  {title} {{Theory and experiment in gravitational physics}}}}\ (\bibinfo
  {publisher} {Cambridge University Press},\ \bibinfo {year}
  {1993})\BibitemShut {NoStop}%
\bibitem [{\citenamefont {Will}(2014)}]{Will:2014kxa}%
  \BibitemOpen
  \bibfield  {author} {\bibinfo {author} {\bibfnamefont {C.~M.}\ \bibnamefont
  {Will}},\ }\bibfield  {title} {\bibinfo {title} {{The Confrontation between
  General Relativity and Experiment}},\ }\href
  {https://doi.org/10.12942/lrr-2014-4} {\bibfield  {journal} {\bibinfo
  {journal} {Living Rev. Rel.}\ }\textbf {\bibinfo {volume} {17}},\ \bibinfo
  {pages} {4} (\bibinfo {year} {2014})},\ \Eprint
  {https://arxiv.org/abs/1403.7377} {arXiv:1403.7377 [gr-qc]} \BibitemShut
  {NoStop}%
\bibitem [{\citenamefont {Will}(2018)}]{Will:2018bme}%
  \BibitemOpen
  \bibfield  {author} {\bibinfo {author} {\bibfnamefont {C.~M.}\ \bibnamefont
  {Will}},\ }\href
  {https://www.cambridge.org/academic/subjects/physics/cosmology-relativity-and-gravitation/theory-and-experiment-gravitational-physics-2nd-edition?format=AR&isbn=9781108679824}
  {\emph {\bibinfo {title} {{Theory and Experiment in Gravitational
  Physics}}}}\ (\bibinfo  {publisher} {Cambridge University Press},\ \bibinfo
  {year} {2018})\BibitemShut {NoStop}%
\bibitem [{\citenamefont {Bertotti}\ \emph {et~al.}(2003)\citenamefont
  {Bertotti}, \citenamefont {Iess},\ and\ \citenamefont
  {Tortora}}]{Bertotti:2003rm}%
  \BibitemOpen
  \bibfield  {author} {\bibinfo {author} {\bibfnamefont {B.}~\bibnamefont
  {Bertotti}}, \bibinfo {author} {\bibfnamefont {L.}~\bibnamefont {Iess}},\
  and\ \bibinfo {author} {\bibfnamefont {P.}~\bibnamefont {Tortora}},\
  }\bibfield  {title} {\bibinfo {title} {{A test of general relativity using
  radio links with the Cassini spacecraft}},\ }\href
  {https://doi.org/10.1038/nature01997} {\bibfield  {journal} {\bibinfo
  {journal} {Nature}\ }\textbf {\bibinfo {volume} {425}},\ \bibinfo {pages}
  {374} (\bibinfo {year} {2003})}\BibitemShut {NoStop}%
\bibitem [{\citenamefont {Fienga}\ \emph {et~al.}(2011)\citenamefont {Fienga},
  \citenamefont {Laskar}, \citenamefont {Kuchynka}, \citenamefont {Manche},
  \citenamefont {Desvignes}, \citenamefont {Gastineau}, \citenamefont
  {Cognard},\ and\ \citenamefont {Theureau}}]{Fienga:2011qh}%
  \BibitemOpen
  \bibfield  {author} {\bibinfo {author} {\bibfnamefont {A.}~\bibnamefont
  {Fienga}}, \bibinfo {author} {\bibfnamefont {J.}~\bibnamefont {Laskar}},
  \bibinfo {author} {\bibfnamefont {P.}~\bibnamefont {Kuchynka}}, \bibinfo
  {author} {\bibfnamefont {H.}~\bibnamefont {Manche}}, \bibinfo {author}
  {\bibfnamefont {G.}~\bibnamefont {Desvignes}}, \bibinfo {author}
  {\bibfnamefont {M.}~\bibnamefont {Gastineau}}, \bibinfo {author}
  {\bibfnamefont {I.}~\bibnamefont {Cognard}},\ and\ \bibinfo {author}
  {\bibfnamefont {G.}~\bibnamefont {Theureau}},\ }\bibfield  {title} {\bibinfo
  {title} {{The INPOP10a planetary ephemeris and its applications in
  fundamental physics}},\ }\href {https://doi.org/10.1007/s10569-011-9377-8}
  {\bibfield  {journal} {\bibinfo  {journal} {Celest. Mech. Dyn. Astron.}\
  }\textbf {\bibinfo {volume} {111}},\ \bibinfo {pages} {363} (\bibinfo {year}
  {2011})},\ \Eprint {https://arxiv.org/abs/1108.5546} {arXiv:1108.5546
  [astro-ph.EP]} \BibitemShut {NoStop}%
\bibitem [{\citenamefont {Fienga}\ \emph {et~al.}(2015)\citenamefont {Fienga},
  \citenamefont {Laskar}, \citenamefont {Exertier}, \citenamefont {Manche},\
  and\ \citenamefont {Gastineau}}]{Fienga:2014bvy}%
  \BibitemOpen
  \bibfield  {author} {\bibinfo {author} {\bibfnamefont {A.}~\bibnamefont
  {Fienga}}, \bibinfo {author} {\bibfnamefont {J.}~\bibnamefont {Laskar}},
  \bibinfo {author} {\bibfnamefont {P.}~\bibnamefont {Exertier}}, \bibinfo
  {author} {\bibfnamefont {H.}~\bibnamefont {Manche}},\ and\ \bibinfo {author}
  {\bibfnamefont {M.}~\bibnamefont {Gastineau}},\ }\bibfield  {title} {\bibinfo
  {title} {{Numerical estimation of the sensitivity of INPOP planetary
  ephemerides to general relativity parameters}},\ }\href
  {https://doi.org/10.1007/s10569-015-9639-y} {\bibfield  {journal} {\bibinfo
  {journal} {Celest. Mech. Dyn. Astron.}\ }\textbf {\bibinfo {volume} {123}},\
  \bibinfo {pages} {325} (\bibinfo {year} {2015})},\ \Eprint
  {https://arxiv.org/abs/1409.4932} {arXiv:1409.4932 [astro-ph.EP]}
  \BibitemShut {NoStop}%
\bibitem [{\citenamefont {Verma}\ \emph {et~al.}(2014)\citenamefont {Verma},
  \citenamefont {Fienga}, \citenamefont {Laskar}, \citenamefont {Manche},\ and\
  \citenamefont {Gastineau}}]{Verma:2013ata}%
  \BibitemOpen
  \bibfield  {author} {\bibinfo {author} {\bibfnamefont {A.}~\bibnamefont
  {Verma}}, \bibinfo {author} {\bibfnamefont {A.}~\bibnamefont {Fienga}},
  \bibinfo {author} {\bibfnamefont {J.}~\bibnamefont {Laskar}}, \bibinfo
  {author} {\bibfnamefont {H.}~\bibnamefont {Manche}},\ and\ \bibinfo {author}
  {\bibfnamefont {M.}~\bibnamefont {Gastineau}},\ }\bibfield  {title} {\bibinfo
  {title} {{Use of MESSENGER radioscience data to improve planetary and to test
  general relativity}},\ }\href {https://doi.org/10.1051/0004-6361/201322124}
  {\bibfield  {journal} {\bibinfo  {journal} {Astron. Astrophys.}\ }\textbf
  {\bibinfo {volume} {561}},\ \bibinfo {pages} {A115} (\bibinfo {year}
  {2014})},\ \Eprint {https://arxiv.org/abs/1306.5569} {arXiv:1306.5569
  [astro-ph.EP]} \BibitemShut {NoStop}%
\bibitem [{\citenamefont {Nordtvedt}(1970)}]{Nordtvedt:1970uv}%
  \BibitemOpen
  \bibfield  {author} {\bibinfo {author} {\bibfnamefont {K.}~\bibnamefont
  {Nordtvedt}, \bibfnamefont {Jr.}},\ }\bibfield  {title} {\bibinfo {title}
  {{Post-Newtonian metric for a general class of scalar tensor gravitational
  theories and observational consequences}},\ }\href
  {https://doi.org/10.1086/150607} {\bibfield  {journal} {\bibinfo  {journal}
  {Astrophys. J.}\ }\textbf {\bibinfo {volume} {161}},\ \bibinfo {pages} {1059}
  (\bibinfo {year} {1970})}\BibitemShut {NoStop}%
\bibitem [{\citenamefont {Gladchenko}\ \emph {et~al.}(1990)\citenamefont
  {Gladchenko}, \citenamefont {Ponomarev},\ and\ \citenamefont
  {Zhytnikov}}]{Gladchenko:1990nw}%
  \BibitemOpen
  \bibfield  {author} {\bibinfo {author} {\bibfnamefont {M.~S.}\ \bibnamefont
  {Gladchenko}}, \bibinfo {author} {\bibfnamefont {V.~N.}\ \bibnamefont
  {Ponomarev}},\ and\ \bibinfo {author} {\bibfnamefont {V.~V.}\ \bibnamefont
  {Zhytnikov}},\ }\bibfield  {title} {\bibinfo {title} {{PPN metric and PPN
  torsion in the quadratic Poincare gauge theory of gravity}},\ }\href
  {https://doi.org/10.1016/0370-2693(90)91488-W} {\bibfield  {journal}
  {\bibinfo  {journal} {Phys. Lett.}\ }\textbf {\bibinfo {volume} {B241}},\
  \bibinfo {pages} {67} (\bibinfo {year} {1990})}\BibitemShut {NoStop}%
\bibitem [{\citenamefont {Olmo}(2005)}]{Olmo:2005hc}%
  \BibitemOpen
  \bibfield  {author} {\bibinfo {author} {\bibfnamefont {G.~J.}\ \bibnamefont
  {Olmo}},\ }\bibfield  {title} {\bibinfo {title} {{Post-Newtonian constraints
  on f(R) cosmologies in metric and Palatini formalism}},\ }\href
  {https://doi.org/10.1103/PhysRevD.72.083505} {\bibfield  {journal} {\bibinfo
  {journal} {Phys. Rev.}\ }\textbf {\bibinfo {volume} {D72}},\ \bibinfo {pages}
  {083505} (\bibinfo {year} {2005})},\ \Eprint
  {https://arxiv.org/abs/gr-qc/0505135} {arXiv:gr-qc/0505135 [gr-qc]}
  \BibitemShut {NoStop}%
\bibitem [{\citenamefont {Clifton}\ \emph {et~al.}(2010)\citenamefont
  {Clifton}, \citenamefont {Banados},\ and\ \citenamefont
  {Skordis}}]{Clifton:2010hz}%
  \BibitemOpen
  \bibfield  {author} {\bibinfo {author} {\bibfnamefont {T.}~\bibnamefont
  {Clifton}}, \bibinfo {author} {\bibfnamefont {M.}~\bibnamefont {Banados}},\
  and\ \bibinfo {author} {\bibfnamefont {C.}~\bibnamefont {Skordis}},\
  }\bibfield  {title} {\bibinfo {title} {{The Parameterised Post-Newtonian
  Limit of Bimetric Theories of Gravity}},\ }\href
  {https://doi.org/10.1088/0264-9381/27/23/235020} {\bibfield  {journal}
  {\bibinfo  {journal} {Class. Quant. Grav.}\ }\textbf {\bibinfo {volume}
  {27}},\ \bibinfo {pages} {235020} (\bibinfo {year} {2010})},\ \Eprint
  {https://arxiv.org/abs/1006.5619} {arXiv:1006.5619 [gr-qc]} \BibitemShut
  {NoStop}%
\bibitem [{\citenamefont {Perivolaropoulos}(2010)}]{Perivolaropoulos:2009ak}%
  \BibitemOpen
  \bibfield  {author} {\bibinfo {author} {\bibfnamefont {L.}~\bibnamefont
  {Perivolaropoulos}},\ }\bibfield  {title} {\bibinfo {title} {{PPN Parameter
  gamma and Solar System Constraints of Massive Brans-Dicke Theories}},\ }\href
  {https://doi.org/10.1103/PhysRevD.81.047501} {\bibfield  {journal} {\bibinfo
  {journal} {Phys. Rev.}\ }\textbf {\bibinfo {volume} {D81}},\ \bibinfo {pages}
  {047501} (\bibinfo {year} {2010})},\ \Eprint
  {https://arxiv.org/abs/0911.3401} {arXiv:0911.3401 [gr-qc]} \BibitemShut
  {NoStop}%
\bibitem [{\citenamefont {Schärer}\ \emph {et~al.}(2014)\citenamefont
  {Schärer}, \citenamefont {Angélil}, \citenamefont {Bondarescu},
  \citenamefont {Jetzer},\ and\ \citenamefont {Lundgren}}]{Scharer:2014kya}%
  \BibitemOpen
  \bibfield  {author} {\bibinfo {author} {\bibfnamefont {A.}~\bibnamefont
  {Schärer}}, \bibinfo {author} {\bibfnamefont {R.}~\bibnamefont {Angélil}},
  \bibinfo {author} {\bibfnamefont {R.}~\bibnamefont {Bondarescu}}, \bibinfo
  {author} {\bibfnamefont {P.}~\bibnamefont {Jetzer}},\ and\ \bibinfo {author}
  {\bibfnamefont {A.}~\bibnamefont {Lundgren}},\ }\bibfield  {title} {\bibinfo
  {title} {{Testing scalar-tensor theories and parametrized post-Newtonian
  parameters in Earth orbit}},\ }\href
  {https://doi.org/10.1103/PhysRevD.90.123005} {\bibfield  {journal} {\bibinfo
  {journal} {Phys. Rev.}\ }\textbf {\bibinfo {volume} {D90}},\ \bibinfo {pages}
  {123005} (\bibinfo {year} {2014})},\ \Eprint
  {https://arxiv.org/abs/1410.7914} {arXiv:1410.7914 [gr-qc]} \BibitemShut
  {NoStop}%
\bibitem [{\citenamefont {Hohmann}(2014)}]{Hohmann:2013oca}%
  \BibitemOpen
  \bibfield  {author} {\bibinfo {author} {\bibfnamefont {M.}~\bibnamefont
  {Hohmann}},\ }\bibfield  {title} {\bibinfo {title} {{Parameterized
  post-Newtonian formalism for multimetric gravity}},\ }\href
  {https://doi.org/10.1088/0264-9381/31/13/135003} {\bibfield  {journal}
  {\bibinfo  {journal} {Class. Quant. Grav.}\ }\textbf {\bibinfo {volume}
  {31}},\ \bibinfo {pages} {135003} (\bibinfo {year} {2014})},\ \Eprint
  {https://arxiv.org/abs/1309.7787} {arXiv:1309.7787 [gr-qc]} \BibitemShut
  {NoStop}%
\bibitem [{\citenamefont {Li}\ \emph {et~al.}(2014)\citenamefont {Li},
  \citenamefont {Wu},\ and\ \citenamefont {Geng}}]{Li:2013oef}%
  \BibitemOpen
  \bibfield  {author} {\bibinfo {author} {\bibfnamefont {J.-T.}\ \bibnamefont
  {Li}}, \bibinfo {author} {\bibfnamefont {Y.-P.}\ \bibnamefont {Wu}},\ and\
  \bibinfo {author} {\bibfnamefont {C.-Q.}\ \bibnamefont {Geng}},\ }\bibfield
  {title} {\bibinfo {title} {{Parametrized post-Newtonian limit of the
  teleparallel dark energy model}},\ }\href
  {https://doi.org/10.1103/PhysRevD.89.044040} {\bibfield  {journal} {\bibinfo
  {journal} {Phys. Rev.}\ }\textbf {\bibinfo {volume} {D89}},\ \bibinfo {pages}
  {044040} (\bibinfo {year} {2014})},\ \Eprint
  {https://arxiv.org/abs/1312.4332} {arXiv:1312.4332 [gr-qc]} \BibitemShut
  {NoStop}%
\bibitem [{\citenamefont {Hohmann}(2015)}]{Hohmann:2015kra}%
  \BibitemOpen
  \bibfield  {author} {\bibinfo {author} {\bibfnamefont {M.}~\bibnamefont
  {Hohmann}},\ }\bibfield  {title} {\bibinfo {title} {{Parametrized
  post-Newtonian limit of Horndeski’s gravity theory}},\ }\href
  {https://doi.org/10.1103/PhysRevD.92.064019} {\bibfield  {journal} {\bibinfo
  {journal} {Phys. Rev.}\ }\textbf {\bibinfo {volume} {D92}},\ \bibinfo {pages}
  {064019} (\bibinfo {year} {2015})},\ \Eprint
  {https://arxiv.org/abs/1506.04253} {arXiv:1506.04253 [gr-qc]} \BibitemShut
  {NoStop}%
\bibitem [{\citenamefont {Hohmann}\ \emph {et~al.}(2016)\citenamefont
  {Hohmann}, \citenamefont {Jarv}, \citenamefont {Kuusk}, \citenamefont
  {Randla},\ and\ \citenamefont {Vilson}}]{Hohmann:2016yfd}%
  \BibitemOpen
  \bibfield  {author} {\bibinfo {author} {\bibfnamefont {M.}~\bibnamefont
  {Hohmann}}, \bibinfo {author} {\bibfnamefont {L.}~\bibnamefont {Jarv}},
  \bibinfo {author} {\bibfnamefont {P.}~\bibnamefont {Kuusk}}, \bibinfo
  {author} {\bibfnamefont {E.}~\bibnamefont {Randla}},\ and\ \bibinfo {author}
  {\bibfnamefont {O.}~\bibnamefont {Vilson}},\ }\bibfield  {title} {\bibinfo
  {title} {{Post-Newtonian parameter $\gamma$ for multiscalar-tensor gravity
  with a general potential}},\ }\href
  {https://doi.org/10.1103/PhysRevD.94.124015} {\bibfield  {journal} {\bibinfo
  {journal} {Phys. Rev.}\ }\textbf {\bibinfo {volume} {D94}},\ \bibinfo {pages}
  {124015} (\bibinfo {year} {2016})},\ \Eprint
  {https://arxiv.org/abs/1607.02356} {arXiv:1607.02356 [gr-qc]} \BibitemShut
  {NoStop}%
\bibitem [{\citenamefont {Hohmann}(2017)}]{Hohmann:2017uxe}%
  \BibitemOpen
  \bibfield  {author} {\bibinfo {author} {\bibfnamefont {M.}~\bibnamefont
  {Hohmann}},\ }\bibfield  {title} {\bibinfo {title} {{Post-Newtonian parameter
  $\gamma$ and the deflection of light in ghost-free massive bimetric
  gravity}},\ }\href {https://doi.org/10.1103/PhysRevD.95.124049} {\bibfield
  {journal} {\bibinfo  {journal} {Phys. Rev.}\ }\textbf {\bibinfo {volume}
  {D95}},\ \bibinfo {pages} {124049} (\bibinfo {year} {2017})},\ \Eprint
  {https://arxiv.org/abs/1701.07700} {arXiv:1701.07700 [gr-qc]} \BibitemShut
  {NoStop}%
\bibitem [{\citenamefont {Mohseni~Sadjadi}(2017)}]{Sadjadi:2016kwj}%
  \BibitemOpen
  \bibfield  {author} {\bibinfo {author} {\bibfnamefont {H.}~\bibnamefont
  {Mohseni~Sadjadi}},\ }\bibfield  {title} {\bibinfo {title} {{Parameterized
  post-Newtonian approximation in a teleparallel model of dark energy with a
  boundary term}},\ }\href {https://doi.org/10.1140/epjc/s10052-017-4760-6}
  {\bibfield  {journal} {\bibinfo  {journal} {Eur. Phys. J.}\ }\textbf
  {\bibinfo {volume} {C77}},\ \bibinfo {pages} {191} (\bibinfo {year}
  {2017})},\ \Eprint {https://arxiv.org/abs/1606.04362} {arXiv:1606.04362
  [gr-qc]} \BibitemShut {NoStop}%
\bibitem [{\citenamefont {Hohmann}(2020)}]{Hohmann:2019qgo}%
  \BibitemOpen
  \bibfield  {author} {\bibinfo {author} {\bibfnamefont {M.}~\bibnamefont
  {Hohmann}},\ }\bibfield  {title} {\bibinfo {title} {{Gauge-invariant approach
  to the parametrized post-Newtonian formalism}},\ }\href
  {https://doi.org/10.1103/PhysRevD.101.024061} {\bibfield  {journal} {\bibinfo
   {journal} {Phys. Rev.}\ }\textbf {\bibinfo {volume} {D101}},\ \bibinfo
  {pages} {024061} (\bibinfo {year} {2020})},\ \Eprint
  {https://arxiv.org/abs/1910.09245} {arXiv:1910.09245 [gr-qc]} \BibitemShut
  {NoStop}%
\bibitem [{\citenamefont {Ualikhanova}\ and\ \citenamefont
  {Hohmann}(2019)}]{Ualikhanova:2019ygl}%
  \BibitemOpen
  \bibfield  {author} {\bibinfo {author} {\bibfnamefont {U.}~\bibnamefont
  {Ualikhanova}}\ and\ \bibinfo {author} {\bibfnamefont {M.}~\bibnamefont
  {Hohmann}},\ }\bibfield  {title} {\bibinfo {title} {{Parameterized
  post-Newtonian limit of general teleparallel gravity theories}},\ }\href
  {https://doi.org/10.1103/PhysRevD.100.104011} {\bibfield  {journal} {\bibinfo
   {journal} {Phys. Rev.}\ }\textbf {\bibinfo {volume} {D100}},\ \bibinfo
  {pages} {104011} (\bibinfo {year} {2019})},\ \Eprint
  {https://arxiv.org/abs/1907.08178} {arXiv:1907.08178 [gr-qc]} \BibitemShut
  {NoStop}%
\bibitem [{\citenamefont {Emtsova}\ and\ \citenamefont
  {Hohmann}(2020)}]{Emtsova:2019qsl}%
  \BibitemOpen
  \bibfield  {author} {\bibinfo {author} {\bibfnamefont {E.~D.}\ \bibnamefont
  {Emtsova}}\ and\ \bibinfo {author} {\bibfnamefont {M.}~\bibnamefont
  {Hohmann}},\ }\bibfield  {title} {\bibinfo {title} {{Post-Newtonian limit of
  scalar-torsion theories of gravity as analogue to scalar-curvature
  theories}},\ }\href {https://doi.org/10.1103/PhysRevD.101.024017} {\bibfield
  {journal} {\bibinfo  {journal} {Phys. Rev.}\ }\textbf {\bibinfo {volume}
  {D101}},\ \bibinfo {pages} {024017} (\bibinfo {year} {2020})},\ \Eprint
  {https://arxiv.org/abs/1909.09355} {arXiv:1909.09355 [gr-qc]} \BibitemShut
  {NoStop}%
\bibitem [{\citenamefont {Flathmann}\ and\ \citenamefont
  {Hohmann}(2020)}]{Flathmann:2019khc}%
  \BibitemOpen
  \bibfield  {author} {\bibinfo {author} {\bibfnamefont {K.}~\bibnamefont
  {Flathmann}}\ and\ \bibinfo {author} {\bibfnamefont {M.}~\bibnamefont
  {Hohmann}},\ }\bibfield  {title} {\bibinfo {title} {{Post-Newtonian Limit of
  Generalized Scalar-Torsion Theories of Gravity}},\ }\href
  {https://doi.org/10.1103/PhysRevD.101.024005} {\bibfield  {journal} {\bibinfo
   {journal} {Phys. Rev.}\ }\textbf {\bibinfo {volume} {D101}},\ \bibinfo
  {pages} {024005} (\bibinfo {year} {2020})},\ \Eprint
  {https://arxiv.org/abs/1910.01023} {arXiv:1910.01023 [gr-qc]} \BibitemShut
  {NoStop}%
\bibitem [{\citenamefont {Flathmann}\ and\ \citenamefont
  {Hohmann}(2021)}]{Flathmann:2021khc}%
  \BibitemOpen
  \bibfield  {author} {\bibinfo {author} {\bibfnamefont {K.}~\bibnamefont
  {Flathmann}}\ and\ \bibinfo {author} {\bibfnamefont {M.}~\bibnamefont
  {Hohmann}},\ }\bibfield  {title} {\bibinfo {title} {{Post-Newtonian limit of
  generalized symmetric teleparallel gravity}},\ }\href
  {https://doi.org/10.1103/PhysRevD.103.044030} {\bibfield  {journal} {\bibinfo
   {journal} {Phys. Rev.}\ }\textbf {\bibinfo {volume} {D103}},\ \bibinfo
  {pages} {044030} (\bibinfo {year} {2021})},\ \Eprint
  {https://arxiv.org/abs/2012.12875} {arXiv:2012.12875 [gr-qc]} \BibitemShut
  {NoStop}%
\bibitem [{\citenamefont {Hohmann}(2021)}]{Hohmann:2021fpr}%
  \BibitemOpen
  \bibfield  {author} {\bibinfo {author} {\bibfnamefont {M.}~\bibnamefont
  {Hohmann}},\ }\bibfield  {title} {\bibinfo {title} {{Variational Principles
  in Teleparallel Gravity Theories}},\ }\href
  {https://doi.org/10.3390/universe7050114} {\bibfield  {journal} {\bibinfo
  {journal} {Universe}\ }\textbf {\bibinfo {volume} {7}},\ \bibinfo {pages}
  {114} (\bibinfo {year} {2021})},\ \Eprint {https://arxiv.org/abs/2104.00536}
  {arXiv:2104.00536 [gr-qc]} \BibitemShut {NoStop}%
\bibitem [{\citenamefont {Faraoni}(2004)}]{Faraoni:2004pi}%
  \BibitemOpen
  \bibfield  {author} {\bibinfo {author} {\bibfnamefont {V.}~\bibnamefont
  {Faraoni}},\ }\href {https://doi.org/10.1007/978-1-4020-1989-0} {\emph
  {\bibinfo {title} {{Cosmology in scalar tensor gravity}}}},\ Vol.\ \bibinfo
  {volume} {139}\ (\bibinfo {year} {2004})\BibitemShut {NoStop}%
\bibitem [{\citenamefont {Fujii}\ and\ \citenamefont
  {Maeda}(2007)}]{Fujii:2003pa}%
  \BibitemOpen
  \bibfield  {author} {\bibinfo {author} {\bibfnamefont {Y.}~\bibnamefont
  {Fujii}}\ and\ \bibinfo {author} {\bibfnamefont {K.}~\bibnamefont {Maeda}},\
  }\href {http://www.cambridge.org/uk/catalogue/catalogue.asp?isbn=0521811597}
  {\emph {\bibinfo {title} {{The scalar-tensor theory of gravitation}}}}\
  (\bibinfo  {publisher} {Cambridge University Press},\ \bibinfo {year}
  {2007})\BibitemShut {NoStop}%
\bibitem [{\citenamefont {Hohmann}\ \emph {et~al.}(2018)\citenamefont
  {Hohmann}, \citenamefont {Järv},\ and\ \citenamefont
  {Ualikhanova}}]{Hohmann:2018rwf}%
  \BibitemOpen
  \bibfield  {author} {\bibinfo {author} {\bibfnamefont {M.}~\bibnamefont
  {Hohmann}}, \bibinfo {author} {\bibfnamefont {L.}~\bibnamefont {Järv}},\
  and\ \bibinfo {author} {\bibfnamefont {U.}~\bibnamefont {Ualikhanova}},\
  }\bibfield  {title} {\bibinfo {title} {{Covariant formulation of
  scalar-torsion gravity}},\ }\href
  {https://doi.org/10.1103/PhysRevD.97.104011} {\bibfield  {journal} {\bibinfo
  {journal} {Phys. Rev.}\ }\textbf {\bibinfo {volume} {D97}},\ \bibinfo {pages}
  {104011} (\bibinfo {year} {2018})},\ \Eprint
  {https://arxiv.org/abs/1801.05786} {arXiv:1801.05786 [gr-qc]} \BibitemShut
  {NoStop}%
\bibitem [{\citenamefont {Deng}\ and\ \citenamefont
  {Xie}(2016)}]{Deng:2016moh}%
  \BibitemOpen
  \bibfield  {author} {\bibinfo {author} {\bibfnamefont {X.-M.}\ \bibnamefont
  {Deng}}\ and\ \bibinfo {author} {\bibfnamefont {Y.}~\bibnamefont {Xie}},\
  }\bibfield  {title} {\bibinfo {title} {{Solar System tests of a scalar-tensor
  gravity with a general potential: Insensitivity of light deflection and
  Cassini tracking}},\ }\href {https://doi.org/10.1103/PhysRevD.93.044013}
  {\bibfield  {journal} {\bibinfo  {journal} {Phys. Rev. D}\ }\textbf {\bibinfo
  {volume} {93}},\ \bibinfo {pages} {044013} (\bibinfo {year}
  {2016})}\BibitemShut {NoStop}%
\bibitem [{\citenamefont {Hohmann}\ and\ \citenamefont
  {Schärer}(2017)}]{Hohmann:2017qje}%
  \BibitemOpen
  \bibfield  {author} {\bibinfo {author} {\bibfnamefont {M.}~\bibnamefont
  {Hohmann}}\ and\ \bibinfo {author} {\bibfnamefont {A.}~\bibnamefont
  {Schärer}},\ }\bibfield  {title} {\bibinfo {title} {{Post-Newtonian
  parameters $\gamma$ and $\beta$ of scalar-tensor gravity for a homogeneous
  gravitating sphere}},\ }\href {https://doi.org/10.1103/PhysRevD.96.104026}
  {\bibfield  {journal} {\bibinfo  {journal} {Phys. Rev.}\ }\textbf {\bibinfo
  {volume} {D96}},\ \bibinfo {pages} {104026} (\bibinfo {year} {2017})},\
  \Eprint {https://arxiv.org/abs/1708.07851} {arXiv:1708.07851 [gr-qc]}
  \BibitemShut {NoStop}%
\bibitem [{\citenamefont {Golovnev}\ and\ \citenamefont
  {Koivisto}(2018)}]{Golovnev:2018wbh}%
  \BibitemOpen
  \bibfield  {author} {\bibinfo {author} {\bibfnamefont {A.}~\bibnamefont
  {Golovnev}}\ and\ \bibinfo {author} {\bibfnamefont {T.}~\bibnamefont
  {Koivisto}},\ }\bibfield  {title} {\bibinfo {title} {{Cosmological
  perturbations in modified teleparallel gravity models}},\ }\href
  {https://doi.org/10.1088/1475-7516/2018/11/012} {\bibfield  {journal}
  {\bibinfo  {journal} {JCAP}\ }\textbf {\bibinfo {volume} {1811}}\bibfield
  {number} {\bibinfo  {number} { (11)},\ \bibinfo {pages} {012}},\ }\Eprint
  {https://arxiv.org/abs/1808.05565} {arXiv:1808.05565 [gr-qc]} \BibitemShut
  {NoStop}%
\bibitem [{\citenamefont {Golovnev}\ and\ \citenamefont
  {Guzm\'an}(2021)}]{Golovnev:2020zpv}%
  \BibitemOpen
  \bibfield  {author} {\bibinfo {author} {\bibfnamefont {A.}~\bibnamefont
  {Golovnev}}\ and\ \bibinfo {author} {\bibfnamefont {M.-J.}\ \bibnamefont
  {Guzm\'an}},\ }\bibfield  {title} {\bibinfo {title} {{Foundational issues in
  f(T) gravity theory}},\ }\href {https://doi.org/10.1142/S0219887821400077}
  {\bibfield  {journal} {\bibinfo  {journal} {Int. J. Geom. Meth. Mod. Phys.}\
  }\textbf {\bibinfo {volume} {18}},\ \bibinfo {pages} {2140007} (\bibinfo
  {year} {2021})},\ \Eprint {https://arxiv.org/abs/2012.14408}
  {arXiv:2012.14408 [gr-qc]} \BibitemShut {NoStop}%
\bibitem [{\citenamefont {Beltr\'an~Jim\'enez}\ \emph
  {et~al.}(2021)\citenamefont {Beltr\'an~Jim\'enez}, \citenamefont {Golovnev},
  \citenamefont {Koivisto},\ and\ \citenamefont
  {Veerm\"ae}}]{BeltranJimenez:2020fvy}%
  \BibitemOpen
  \bibfield  {author} {\bibinfo {author} {\bibfnamefont {J.}~\bibnamefont
  {Beltr\'an~Jim\'enez}}, \bibinfo {author} {\bibfnamefont {A.}~\bibnamefont
  {Golovnev}}, \bibinfo {author} {\bibfnamefont {T.}~\bibnamefont {Koivisto}},\
  and\ \bibinfo {author} {\bibfnamefont {H.}~\bibnamefont {Veerm\"ae}},\
  }\bibfield  {title} {\bibinfo {title} {{Minkowski space in $f(T)$ gravity}},\
  }\href {https://doi.org/10.1103/PhysRevD.103.024054} {\bibfield  {journal}
  {\bibinfo  {journal} {Phys. Rev. D}\ }\textbf {\bibinfo {volume} {103}},\
  \bibinfo {pages} {024054} (\bibinfo {year} {2021})},\ \Eprint
  {https://arxiv.org/abs/2004.07536} {arXiv:2004.07536 [gr-qc]} \BibitemShut
  {NoStop}%
\bibitem [{\citenamefont {Blagojevi\'c}\ and\ \citenamefont
  {Nester}(2020)}]{Blagojevic:2020dyq}%
  \BibitemOpen
  \bibfield  {author} {\bibinfo {author} {\bibfnamefont {M.}~\bibnamefont
  {Blagojevi\'c}}\ and\ \bibinfo {author} {\bibfnamefont {J.~M.}\ \bibnamefont
  {Nester}},\ }\bibfield  {title} {\bibinfo {title} {{Local symmetries and
  physical degrees of freedom in $f(T)$ gravity: a Dirac Hamiltonian constraint
  analysis}},\ }\href {https://doi.org/10.1103/PhysRevD.102.064025} {\bibfield
  {journal} {\bibinfo  {journal} {Phys. Rev. D}\ }\textbf {\bibinfo {volume}
  {102}},\ \bibinfo {pages} {064025} (\bibinfo {year} {2020})},\ \Eprint
  {https://arxiv.org/abs/2006.15303} {arXiv:2006.15303 [gr-qc]} \BibitemShut
  {NoStop}%
\bibitem [{\citenamefont {Guzm\'an}\ and\ \citenamefont
  {Ferraro}(2020)}]{Guzman:2019oth}%
  \BibitemOpen
  \bibfield  {author} {\bibinfo {author} {\bibfnamefont {M.~J.}\ \bibnamefont
  {Guzm\'an}}\ and\ \bibinfo {author} {\bibfnamefont {R.}~\bibnamefont
  {Ferraro}},\ }\bibfield  {title} {\bibinfo {title} {{Degrees of freedom and
  Hamiltonian formalism for $f(T)$ gravity}},\ }\href
  {https://doi.org/10.1142/S0217751X20400229} {\bibfield  {journal} {\bibinfo
  {journal} {Int. J. Mod. Phys. A}\ }\textbf {\bibinfo {volume} {35}},\
  \bibinfo {pages} {2040022} (\bibinfo {year} {2020})},\ \Eprint
  {https://arxiv.org/abs/1910.03100} {arXiv:1910.03100 [gr-qc]} \BibitemShut
  {NoStop}%
\bibitem [{\citenamefont {Beltr\'an~Jim\'enez}\ and\ \citenamefont
  {Dialektopoulos}(2020)}]{BeltranJimenez:2019nns}%
  \BibitemOpen
  \bibfield  {author} {\bibinfo {author} {\bibfnamefont {J.}~\bibnamefont
  {Beltr\'an~Jim\'enez}}\ and\ \bibinfo {author} {\bibfnamefont {K.~F.}\
  \bibnamefont {Dialektopoulos}},\ }\bibfield  {title} {\bibinfo {title}
  {{Non-Linear Obstructions for Consistent New General Relativity}},\ }\href
  {https://doi.org/10.1088/1475-7516/2020/01/018} {\bibfield  {journal}
  {\bibinfo  {journal} {JCAP}\ }\textbf {\bibinfo {volume} {01}},\ \bibinfo
  {pages} {018}},\ \Eprint {https://arxiv.org/abs/1907.10038} {arXiv:1907.10038
  [gr-qc]} \BibitemShut {NoStop}%
\bibitem [{\citenamefont {Bahamonde}\ \emph
  {et~al.}(2023{\natexlab{a}})\citenamefont {Bahamonde}, \citenamefont
  {Dialektopoulos}, \citenamefont {Hohmann}, \citenamefont {Levi~Said},
  \citenamefont {Pfeifer},\ and\ \citenamefont
  {Saridakis}}]{Bahamonde:2022ohm}%
  \BibitemOpen
  \bibfield  {author} {\bibinfo {author} {\bibfnamefont {S.}~\bibnamefont
  {Bahamonde}}, \bibinfo {author} {\bibfnamefont {K.~F.}\ \bibnamefont
  {Dialektopoulos}}, \bibinfo {author} {\bibfnamefont {M.}~\bibnamefont
  {Hohmann}}, \bibinfo {author} {\bibfnamefont {J.}~\bibnamefont {Levi~Said}},
  \bibinfo {author} {\bibfnamefont {C.}~\bibnamefont {Pfeifer}},\ and\ \bibinfo
  {author} {\bibfnamefont {E.~N.}\ \bibnamefont {Saridakis}},\ }\bibfield
  {title} {\bibinfo {title} {{Perturbations in non-flat cosmology for f(T)
  gravity}},\ }\href {https://doi.org/10.1140/epjc/s10052-023-11322-3}
  {\bibfield  {journal} {\bibinfo  {journal} {Eur. Phys. J. C}\ }\textbf
  {\bibinfo {volume} {83}},\ \bibinfo {pages} {193} (\bibinfo {year}
  {2023}{\natexlab{a}})},\ \Eprint {https://arxiv.org/abs/2203.00619}
  {arXiv:2203.00619 [gr-qc]} \BibitemShut {NoStop}%
\bibitem [{\citenamefont {Golovnev}\ and\ \citenamefont
  {Guzman}(2021)}]{Golovnev:2020nln}%
  \BibitemOpen
  \bibfield  {author} {\bibinfo {author} {\bibfnamefont {A.}~\bibnamefont
  {Golovnev}}\ and\ \bibinfo {author} {\bibfnamefont {M.-J.}\ \bibnamefont
  {Guzman}},\ }\bibfield  {title} {\bibinfo {title} {{Nontrivial Minkowski
  backgrounds in $f(T)$ gravity}},\ }\href
  {https://doi.org/10.1103/PhysRevD.103.044009} {\bibfield  {journal} {\bibinfo
   {journal} {Phys. Rev. D}\ }\textbf {\bibinfo {volume} {103}},\ \bibinfo
  {pages} {044009} (\bibinfo {year} {2021})},\ \Eprint
  {https://arxiv.org/abs/2012.00696} {arXiv:2012.00696 [gr-qc]} \BibitemShut
  {NoStop}%
\bibitem [{\citenamefont {Li}\ and\ \citenamefont {Rao}(2023)}]{Li:2023fto}%
  \BibitemOpen
  \bibfield  {author} {\bibinfo {author} {\bibfnamefont {M.}~\bibnamefont
  {Li}}\ and\ \bibinfo {author} {\bibfnamefont {H.}~\bibnamefont {Rao}},\
  }\bibfield  {title} {\bibinfo {title} {{Irregular universe in the Nieh-Yan
  modified teleparallel gravity}},\ }\href
  {https://doi.org/10.1016/j.physletb.2023.137929} {\bibfield  {journal}
  {\bibinfo  {journal} {Phys. Lett. B}\ }\textbf {\bibinfo {volume} {841}},\
  \bibinfo {pages} {137929} (\bibinfo {year} {2023})},\ \Eprint
  {https://arxiv.org/abs/2301.02847} {arXiv:2301.02847 [gr-qc]} \BibitemShut
  {NoStop}%
\bibitem [{\citenamefont {Horndeski}(1974)}]{Horndeski:1974wa}%
  \BibitemOpen
  \bibfield  {author} {\bibinfo {author} {\bibfnamefont {G.~W.}\ \bibnamefont
  {Horndeski}},\ }\bibfield  {title} {\bibinfo {title} {{Second-order
  scalar-tensor field equations in a four-dimensional space}},\ }\href
  {https://doi.org/10.1007/BF01807638} {\bibfield  {journal} {\bibinfo
  {journal} {Int. J. Theor. Phys.}\ }\textbf {\bibinfo {volume} {10}},\
  \bibinfo {pages} {363} (\bibinfo {year} {1974})}\BibitemShut {NoStop}%
\bibitem [{\citenamefont {Kobayashi}\ \emph {et~al.}(2011)\citenamefont
  {Kobayashi}, \citenamefont {Yamaguchi},\ and\ \citenamefont
  {Yokoyama}}]{Kobayashi:2011nu}%
  \BibitemOpen
  \bibfield  {author} {\bibinfo {author} {\bibfnamefont {T.}~\bibnamefont
  {Kobayashi}}, \bibinfo {author} {\bibfnamefont {M.}~\bibnamefont
  {Yamaguchi}},\ and\ \bibinfo {author} {\bibfnamefont {J.}~\bibnamefont
  {Yokoyama}},\ }\bibfield  {title} {\bibinfo {title} {{Generalized
  G-inflation: Inflation with the most general second-order field equations}},\
  }\href {https://doi.org/10.1143/PTP.126.511} {\bibfield  {journal} {\bibinfo
  {journal} {Prog. Theor. Phys.}\ }\textbf {\bibinfo {volume} {126}},\ \bibinfo
  {pages} {511} (\bibinfo {year} {2011})},\ \Eprint
  {https://arxiv.org/abs/1105.5723} {arXiv:1105.5723 [hep-th]} \BibitemShut
  {NoStop}%
\bibitem [{\citenamefont {Kobayashi}(2019)}]{Kobayashi:2019hrl}%
  \BibitemOpen
  \bibfield  {author} {\bibinfo {author} {\bibfnamefont {T.}~\bibnamefont
  {Kobayashi}},\ }\bibfield  {title} {\bibinfo {title} {{Horndeski theory and
  beyond: a review}},\ }\href {https://doi.org/10.1088/1361-6633/ab2429}
  {\bibfield  {journal} {\bibinfo  {journal} {Rept. Prog. Phys.}\ }\textbf
  {\bibinfo {volume} {82}},\ \bibinfo {pages} {086901} (\bibinfo {year}
  {2019})},\ \Eprint {https://arxiv.org/abs/1901.07183} {arXiv:1901.07183
  [gr-qc]} \BibitemShut {NoStop}%
\bibitem [{\citenamefont {Bahamonde}\ \emph {et~al.}(2019)\citenamefont
  {Bahamonde}, \citenamefont {Dialektopoulos},\ and\ \citenamefont
  {Said}}]{Bahamonde:2019shr}%
  \BibitemOpen
  \bibfield  {author} {\bibinfo {author} {\bibfnamefont {S.}~\bibnamefont
  {Bahamonde}}, \bibinfo {author} {\bibfnamefont {K.~F.}\ \bibnamefont
  {Dialektopoulos}},\ and\ \bibinfo {author} {\bibfnamefont {J.~L.}\
  \bibnamefont {Said}},\ }\bibfield  {title} {\bibinfo {title} {{Can Horndeski
  Theory be recast using Teleparallel Gravity?}},\ }\href
  {https://doi.org/10.1103/PhysRevD.100.064018} {\bibfield  {journal} {\bibinfo
   {journal} {Phys. Rev.}\ }\textbf {\bibinfo {volume} {D100}},\ \bibinfo
  {pages} {064018} (\bibinfo {year} {2019})},\ \Eprint
  {https://arxiv.org/abs/1904.10791} {arXiv:1904.10791 [gr-qc]} \BibitemShut
  {NoStop}%
\bibitem [{\citenamefont {Bahamonde}\ \emph
  {et~al.}(2020{\natexlab{a}})\citenamefont {Bahamonde}, \citenamefont
  {Dialektopoulos}, \citenamefont {Gakis},\ and\ \citenamefont
  {Levi~Said}}]{Bahamonde:2019ipm}%
  \BibitemOpen
  \bibfield  {author} {\bibinfo {author} {\bibfnamefont {S.}~\bibnamefont
  {Bahamonde}}, \bibinfo {author} {\bibfnamefont {K.~F.}\ \bibnamefont
  {Dialektopoulos}}, \bibinfo {author} {\bibfnamefont {V.}~\bibnamefont
  {Gakis}},\ and\ \bibinfo {author} {\bibfnamefont {J.}~\bibnamefont
  {Levi~Said}},\ }\bibfield  {title} {\bibinfo {title} {{Reviving Horndeski
  theory using teleparallel gravity after GW170817}},\ }\href
  {https://doi.org/10.1103/PhysRevD.101.084060} {\bibfield  {journal} {\bibinfo
   {journal} {Phys. Rev. D}\ }\textbf {\bibinfo {volume} {101}},\ \bibinfo
  {pages} {084060} (\bibinfo {year} {2020}{\natexlab{a}})},\ \Eprint
  {https://arxiv.org/abs/1907.10057} {arXiv:1907.10057 [gr-qc]} \BibitemShut
  {NoStop}%
\bibitem [{\citenamefont {Bahamonde}\ \emph
  {et~al.}(2020{\natexlab{b}})\citenamefont {Bahamonde}, \citenamefont
  {Dialektopoulos}, \citenamefont {Hohmann},\ and\ \citenamefont
  {Levi~Said}}]{Bahamonde:2020cfv}%
  \BibitemOpen
  \bibfield  {author} {\bibinfo {author} {\bibfnamefont {S.}~\bibnamefont
  {Bahamonde}}, \bibinfo {author} {\bibfnamefont {K.~F.}\ \bibnamefont
  {Dialektopoulos}}, \bibinfo {author} {\bibfnamefont {M.}~\bibnamefont
  {Hohmann}},\ and\ \bibinfo {author} {\bibfnamefont {J.}~\bibnamefont
  {Levi~Said}},\ }\bibfield  {title} {\bibinfo {title} {{Post-Newtonian limit
  of Teleparallel Horndeski gravity}},\ }\href
  {https://doi.org/10.1088/1361-6382/abc441} {\bibfield  {journal} {\bibinfo
  {journal} {Class. Quant. Grav.}\ }\textbf {\bibinfo {volume} {38}},\ \bibinfo
  {pages} {025006} (\bibinfo {year} {2020}{\natexlab{b}})},\ \Eprint
  {https://arxiv.org/abs/2003.11554} {arXiv:2003.11554 [gr-qc]} \BibitemShut
  {NoStop}%
\bibitem [{\citenamefont {Bahamonde}\ \emph
  {et~al.}(2023{\natexlab{b}})\citenamefont {Bahamonde}, \citenamefont
  {Trenkler}, \citenamefont {Trombetta},\ and\ \citenamefont
  {Yamaguchi}}]{Bahamonde:2022cmz}%
  \BibitemOpen
  \bibfield  {author} {\bibinfo {author} {\bibfnamefont {S.}~\bibnamefont
  {Bahamonde}}, \bibinfo {author} {\bibfnamefont {G.}~\bibnamefont {Trenkler}},
  \bibinfo {author} {\bibfnamefont {L.~G.}\ \bibnamefont {Trombetta}},\ and\
  \bibinfo {author} {\bibfnamefont {M.}~\bibnamefont {Yamaguchi}},\ }\bibfield
  {title} {\bibinfo {title} {{Symmetric teleparallel Horndeski gravity}},\
  }\href {https://doi.org/10.1103/PhysRevD.107.104024} {\bibfield  {journal}
  {\bibinfo  {journal} {Phys. Rev. D}\ }\textbf {\bibinfo {volume} {107}},\
  \bibinfo {pages} {104024} (\bibinfo {year} {2023}{\natexlab{b}})},\ \Eprint
  {https://arxiv.org/abs/2212.08005} {arXiv:2212.08005 [gr-qc]} \BibitemShut
  {NoStop}%
\end{thebibliography}%
\end{document}